\documentclass[AMA,STIX1COL]{WileyNJD-v2}

\usepackage{hyperref}
\hypersetup{
  colorlinks = true, 
  citecolor = blue, 
  linkcolor = blue, 
  urlcolor = black
}
\usepackage{graphicx}
\usepackage{amsmath}
\usepackage{mathtools}
\usepackage{pifont}
\usepackage{xcolor}
\usepackage{float}
\usepackage{url}
\usepackage{tabularx}
\usepackage{slashbox}
\usepackage{algorithm}
\usepackage{algorithmicx}
\usepackage{algpseudocode}
\usepackage{setspace}

\newcommand\StateX{\Statex\hspace{\algorithmicindent}}
\newcommand{\rw}[1]{\textcolor{black}{#1}}

\DeclareMathOperator*{\argmin}{argmin}

\definecolor{grey}{HTML}{757575}

\articletype{RESEARCH ARTICLE}

\received{}
\revised{}
\accepted{}
        
\begin{document}

%%%%%%%%%%%
%% Title %%
%%%%%%%%%%%

\title{PECCO: A Profit and Cost-oriented Computation Offloading Scheme in Edge-Cloud Environment with Improved Moth-flame Optimisation}

%%%%%%%%%%%%%%%%%%%%%%%%
%%% Author & Address %%%
%%%%%%%%%%%%%%%%%%%%%%%%

\author[1,2]{Jiashu Wu}
\author[1,2]{Hao Dai}
\author[1]{Yang Wang*}
\author[3]{Shigen Shen}
\author[4]{Chengzhong Xu}

\authormark{Jiashu Wu \textsc{et al}}

\address[1]{\orgname{Shenzhen Institute of Advanced Technology, Chinese Academy of Sciences}, \orgaddress{\state{Shenzhen 518055}, \country{China}}}
\address[2]{\orgname{University of Chinese Academy of Sciences}, \orgaddress{\state{Beijing 100049}, \country{China}}}
\address[3]{\orgname{Shaoxing University}, \orgaddress{\state{Shaoxing 312000}, \country{China}}}
\address[4]{\orgname{University of Macau}, \orgaddress{\state{Macau 999078}, \country{China}}}

\corres{*Corresponding Author: Yang Wang, \email{yang.wang1@siat.ac.cn}}

\presentaddress{1068 Xueyuan Avenue, Shenzhen University Town, Shenzhen 518055, Guangdong, P.R.China}

%%%%%%%%%%%%%%
%% Abstract %%
%%%%%%%%%%%%%%

\abstract[Summary]{With the fast growing quantity of data generated by smart devices and the exponential surge of processing demand in the Internet of Things (IoT) era, the resource-rich cloud centres have been utilised to tackle these challenges. \rw{To relieve the burden on cloud centres, edge-cloud computation offloading becomes a promising solution since shortening the proximity between the data source and the computation by offloading computation tasks from the cloud to edge devices can improve performance and Quality of Service (QoS). }Several optimisation models of edge-cloud computation offloading have been proposed that take computation costs and heterogeneous communication costs into account. However, several important factors are not jointly considered, such as heterogeneities of tasks, load balancing among nodes and the profit yielded by computation tasks, which lead to the profit and cost-oriented computation offloading optimisation model \emph{PECCO} proposed in this paper. \rw{Considering that the model is hard in nature and the optimisation objective is not differentiable, we propose an improved Moth-flame optimiser \emph{PECCO-MFI} which addresses some deficiencies of the original Moth-flame Optimiser and integrate it under the edge-cloud environment. }Comprehensive experiments are conducted to verify the superior performance of the proposed method when optimising the proposed task offloading model under the edge-cloud environment. }


\maketitle

%%%%%%%%%%%%%%%%%%%%
%%% Introduction %%%
%%%%%%%%%%%%%%%%%%%%

\section{Introduction}
\label{sec:section_introduction}

With the rapid prevalence of smart devices \cite{rapid_prevalence_smart_devices_shafique2020internet} such as mobile phone and Internet-of-Things (IoT) devices \cite{8264678,iot_app,9698094}, a vast amount of data has been generated \cite{wu_theta_join} and the demand of computation resources has been boosted \cite{huge_data_generated_marjani2017big,li2021self_big_data}. Due to the limited computation, storage and energy capacity of these smart devices \cite{iot_limited_computation_shakarami2020survey}, the powerful \emph{cloud computing} has been leveraged to provide elastic on-demand services to cope with limitations of smart devices \cite{9691460,SHEN2022103140}. With the support of resource-rich cloud servers, processing and storage-intensive applications such as Augmented Reality (AR) \cite{ar_ren2019edge} and Virtual Reality (VR) \cite{vr_zhang2017towards} become feasible. 

However, the fast growing of computation demands pose severe burden on cloud centres \cite{burden_on_cloud_centre_mao2017mobile,burden_on_cloud_centre_yu2017survey,burden_on_cloud_centre_shi2016edge}, and tremendous amount of data generated \cite{data_generated,wu_theta_join} congests the network with limited bandwidth \cite{burden_on_bandwidth_shi2016promise,burden_on_cloud_centre_yu2017survey,burden_on_cloud_centre_shi2016edge}, hence causing bottlenecks for the cloud-based computing paradigm. To relieve the pressure on cloud centres, the \emph{edge computing} concept has emerged \cite{edge_computing_survey_khan2019edge,burden_on_cloud_centre_yu2017survey}, which allows computation to be performed at the edge network. The \emph{Edge network} \cite{burden_on_cloud_centre_shi2016edge} refers to the computing and network resources sit along the path between data sources and cloud centres. The rationale of \emph{computation offloading} \cite{heto_du2020algorithmics,computation_offloading_wang2019edge} is to let the computation happen at closer proximity to the data sources, so that not only the load pressure of cloud centres can be lessened, but also the Quality of Service (QoS) can be improved as the edge computing can provide more efficient responses \cite{qos_mach2017mobile}. 

\rw{To fully excavate the potential of edge-cloud computation offloading, several past research efforts investigated performance-influencing factors and proposed optimisation models to maximise the performance gain while not causing significant costs. }Wang et al., \cite{homogeneous_communication_limited_factor_wang2017computational} presented a Two-Phase Optimisation algorithm and an Iterative Improvement algorithm to jointly optimise the computation costs and latency under the mobile-edge setting. Li et al., \cite{limited_factor_li2001computation} constructed a cost graph to optimise the energy consumption of handheld computing devices during computation offloading to achieve considerable energy saving. Works in \cite{homogeneous_communication_limited_factor_wang2017computational,larac_juttner2001lagrange,homogeneous_communication_wu2016optimal,homogeneous_communication_dong2018computation} all considered optimising the communication cost under the edge-cloud task offloading setting. Specifically, they adopted a homogeneous communication model with two assumptions: cloud-cloud and edge-edge communication cost were ignored, and the edge-cloud and cloud-edge communications were assumed to have symmetric costs, irrespective of the communication direction and distance between nodes. To address the over-simplicity of homogeneous communication models in these methods, Du et al., \cite{heto_du2020algorithmics} attempted the heterogeneous communication model and proposed HETO algorithm that can jointly minimise the computation, communication and migration costs during computation offloading. Their work was the first to propose a heterogeneous communication model so that the deficiencies in existing researches can be overcame. 

\rw{Despite various attempts of past researches to optimise the edge-cloud computation offloading problem}, these models still suffered from the following drawbacks: 

\begin{itemize}

  \item \rw{These methods were not fine-grained enough. Although some methods considered heterogeneous communication cost, these methods failed to leverage more fine-grained factors such as the distance between node pairs. These methods also only leveraged a homogeneous cost model for computation tasks, which did not reflect the task heterogeneity in real-world settings. }
  
  \item \rw{During computation offloading, some methods did not pay attention to the load balancing, which can cause overloading on certain nodes. }
  
  \item These methods were cost-oriented, which failed to jointly optimise the profit and cost during computation offloading. 
  
\end{itemize}

To address these issues, we propose a novel edge-cloud computation offloading model which not only utilises the more realistic heterogeneous communication and computation cost model, but also considers the cost and profit heterogeneities of tasks. Hence, the proposed method jointly optimises the profit and cost yielded during computation offloading. The model is named as \emph{PECCO}, which stands for ``\ul{P}rofit and Cost-oriented \ul{E}dge-\ul{C}loud \ul{C}omputation \ul{O}ffloading''. 

Considering this optimisation problem is hard in nature and the objective of the \emph{PECCO} model is not differentiable, we consider using the Moth-flame Optimisation (MFO) algorithm \cite{moth_flame_optimisation_mirjalili2015moth} to tackle the computation offloading problem. \rw{As a swarm-based algorithm, it is gradient-free and it balances exploration and exploitation. Besides, empirically it outperforms other swarm-based counterparts in terms of convergence speed \cite{moth_flame_optimisation_mirjalili2015moth}, etc., which make it suitable to be leveraged in this case. }We therefore propose an improved Moth-flame optimiser (\emph{PECCO-MFI}) that addresses several drawbacks of the original MFO and significantly boosts its optimisation effectiveness. Specifically, a density-aware Moth-flame initialiser is designed to fit under the edge-cloud computation offloading setting. A dynamic hierarchical flaming mechanism is applied to avoid the single flame matching which is more likely to cause local optima stagnation. Moreover, the lifetime of moths is introduced to promote exploration when the corresponding paired flame is eliminated. 

In summary, this paper makes the following contributions: 

\begin{itemize}
  
  \item We construct a profit and cost-oriented edge-cloud computation offloading optimisation model \emph{PECCO} that jointly optimises both the heterogeneous profit of computation tasks and the heterogeneous cost produced during computation offloading. 
  
  \item \rw{We not only utilise the heterogeneous communication cost, but also consider the load balancing among nodes during optimisation. }
  
  \item We realise the suitability to leverage the Moth-flame Optimiser, and propose an improved algorithm which tackles several deficiencies of the original MFO and hence boosts the effectiveness when solving the proposed computation offloading model. 

\end{itemize}

The rest of the paper is organised as follows, Section \ref{sec:section_related_work} introduces some related works on both edge-cloud computation offloading models and optimisation algorithms. The research opportunities are then discussed. \rw{The background, suitability and room for improvements to the Moth-flame Optimisation algorithm are given in Section \ref{sec:section_background}. }Section \ref{sec:section_model_and_method} presents the details of the proposed model, as well as how the Moth-flame Optimisation algorithm is improved and integrated. Section \ref{sec:section_experiment} presents the experimental settings and results to verify the effectiveness of the proposed algorithm when tackling the proposed model. Section \ref{sec:section_conclusion} concludes the paper.

%%%%%%%%%%%%%%%%%%%%
%%% Related Work %%%
%%%%%%%%%%%%%%%%%%%%

\section{Related Work}
\label{sec:section_related_work}

\rw{In this section, past research works on edge-cloud computation offloading models will be presented. Then, some well-known optimisers will be presented and compared. Finally, the research opportunities of our work are discussed. }

\subsection{Offloading Model}
\label{sec:section_edge_cloud_computation_offloading_model_related_works}

As a promising technique that can relieve the burden posed on cloud centres, edge-cloud computation offloading has drawn huge attention from both industry and academic community \cite{iot_limited_computation_shakarami2020survey,9261459}. Wu et al., \cite{homogeneous_communication_wu2016optimal} formulated the edge-cloud computation offloading problem into a graph min-cost partitioning problem, in which computation tasks will be partitioned to be run on either the cloud side or the edge side. The proposed Min-Cost Offloading Partitioning (MCOP) algorithm took both the execution time and energy consumption into account when deciding an optimal task partitioning strategy. Li et al., \cite{limited_factor_li2001computation} put forward a partition scheme to offload computation tasks on handheld devices. A cost graph was constructed and the partition scheme was applied to split computation programs into server tasks and client tasks with the aim to reduce the energy consumption. Juttner et al., \cite{larac_juttner2001lagrange} presented the Lagrange Relaxation based Aggregated Cost (LARAC) algorithm, which formulated a task graph and traversed the shortest path between nodes when considering the communication costs. The proposed algorithm was effective on delay-sensitive applications, justified by the simulation experiment they performed. 

In works completed by Wang et al., \cite{homogeneous_communication_limited_factor_wang2017computational} and Dong et al., \cite{homogeneous_communication_dong2018computation}, they paid attention to the communication cost faced in the edge-cloud computation offloading problem. \rw{When modelling the communication cost, communications between nodes on the same side (cloud-cloud, or edge-edge) were assumed to be cost-free. Moreover, to simplify the model, communication costs were assumed to be symmetric, i.e., cloud-edge and edge-cloud communications have the same cost, irrespective of direction and the distance between nodes on different sides. The homogeneous communication model they leveraged is considered to be over-simplified and highly infeasible in real-world settings, as the cost can be asymmetric and distance-dependent. Therefore, Du et al., \cite{heto_du2020algorithmics} proposed a more fine-grained heterogeneous cost model, in which the symmetric assumption was relaxed, and the communication costs between nodes in a single side were no longer ignored. }They then formulated the problem as a graph partitioning problem and designed the HETO algorithm to find a sub-optimal offloading strategy. Experiments on PageRank datasets testified to the excellent performance of the HETO algorithm when minimising the communication, computation and migration costs. 

Despite that various research efforts have been drawn to optimise the edge-cloud computation offloading, they still suffered from some drawbacks which need to be addressed: 

\begin{itemize}
  
  \item Although the heterogeneous communication cost has been considered in some works, they failed to leverage more fine-grained factors such as distance between node pairs. 
  
  \item When considering the cost during computation offloading, these methods utilised a homogeneous cost model for computation tasks, i.e., task heterogeneity was ignored. 
  
  \item During computation offloading, some methods did not take load balancing into consideration, i.e., some node may be overloaded. 
  
  \item These methods were cost-oriented, which failed to jointly optimise the profit and cost during computation offloading. 

\end{itemize}

\subsection{Model Optimiser}
\label{sec:section_model_optimiser_related_work}

A suitable optimiser is indispensable to tackle the edge-cloud computation offloading problem and find out an excellent offloading strategy. Some well-known individual-based optimisation algorithms were proposed \cite{SHI2018316,lai2020analysis}. They only optimised a single search candidate, and hence enjoyed a lighter computation cost and required less function evaluations. For instance, \rw{Lawrence \cite{hill_climbing_davis1991bit} presented the Hill Climbing (HC) algorithm which iteratively improved a single search candidate by changing its variables. }The Iterated Local Search (ILS) algorithm proposed by Lourenco et al., \cite{ils_lourencco2003iterated} was an improvement towards the HC algorithm. The best solution obtained in each iteration was perturbed and utilised as the starting point of the next iteration. Despite the efficiency enjoyed by these algorithms, they suffered a lot from the local optima stagnation. These algorithms may encounter the premature convergence, which prevents them from converging towards the global optima. Some more advanced algorithms such as gradient descent \cite{gradient_descent_ruder2016overview} have also been widely applied, especially for the optimisation in the field of deep learning \cite{goodfellow2016deep,li2020simultaneous}. \rw{However, these methods required gradient information of the objective function, which made them not applicable when the objective function is not differentiable. }

In order to provide better local optima avoidance, some population-based optimisation algorithms have been proposed and became popular in the past few years. By utilising multiple search candidates and meanwhile balancing between exploration and exploitation, they provided higher possibilities to approach the global optima. \rw{As a sub-class of population-based algorithms, swarm-based algorithms \cite{swarm_keerthi2015survey} utilised multiple search candidates for the purpose of exploration. }These search candidates then iteratively evolve, and eventually the healthier individuals will survive, making the exploitation become possible. \rw{Kennedy et al., \cite{kennedy1995particle} presented the Particle Swarm Optimisation (PSO) algorithm that mimicked the behaviour of birds in a flock which keep track of their individual and global best positions. The PSO involved only primitive math operations and was computationally inexpensive. }Yang \cite{yang2013firefly} proposed the Firefly Optimisation algorithm (FFA), which was inspired by fireflies. During flying, fireflies are attracted by other fireflies with higher brightness. The effectiveness of the algorithm was verified on several test functions. \rw{A Whale Optimisation algorithm (WOA) was proposed by Mirjalili et al., \cite{mirjalili2016whale} which was inspired by the bubble-net hunting strategy of humpback whales. The WOA algorithm mathematically modelled this behaviour to guide optimisation. A Grey Wolf Optimiser (GWO) was also proposed by Mirjalili et al., \cite{2014Grey} which modelled the social hierarchy of grey wolves during hunting to guide the optimisation process. }Besides, Mirjalili \cite{moth_flame_optimisation_mirjalili2015moth} put forward the Moth-flame Optimisation algorithm (MFO), which was one of the most famous swarm-based optimisers. The MFO utilised a population of moths to act as search candidates so that the probability of approximating the global optima was increased. Inspired by the transverse orientation characteristic of moths, the location of moths will be updated based on their transverse oriented path, with extra parameters controlling the exploration and exploitation. Each search candidate was iteratively assessed by a fitness function and hence the MFO algorithm was gradient-free. After several generation of evolvements, the fittest moth will be regarded as the optimised result. \rw{Experiments on several benchmarks \cite{moth_flame_optimisation_mirjalili2015moth} demonstrated that compared with several counterparts, the MFO algorithm can achieve better optimisation results with statistical significance, while also converge in a fast manner. }

\subsection{Research Opportunity}
\label{sec:section_research_opportunity}

Considering that past edge-cloud computation offloading models suffered from these aforementioned drawbacks, we find it promising to propose an optimisation model that is both profit and cost-oriented. In terms of costs, the heterogeneous communication cost should be considered in a fine-grained manner. Moreover, a heterogeneous cost model should also be applied for tasks to make the model more practical. Besides, during computation offloading, load balancing should be taken care of to avoid computation node overloading. 

A comprehensive optimisation model and an excellent optimiser are both indispensable to produce a better task offloading strategy. Given the suitability of the MFO algorithm such as a higher chance to converge towards the global optima and its gradient-free merit, \rw{we propose an improved Moth-flame optimiser that addresses some design flaws of the original MFO, which can boost the effectiveness when working on the proposed computation offloading model. }

%%%%%%%%%%%%%%%%%%
%%% Background %%%
%%%%%%%%%%%%%%%%%%

\section{Background}
\label{sec:section_background}

In this section, we firstly introduce the background of the Moth-flame Optimisation (MFO) algorithm, including how it works, its advantages and its suitability to be utilised to solve the \emph{PECCO} model. Then, some deficiencies of the MFO algorithm are pointed out which \rw{provide room for improvements for its improved version. }

\subsection{The Moth-flame Optimisation Algorithm}
\label{sec:section_the_moth_flame_optimisation_algorithm}

\begin{figure}[!h]
  \begin{center}
    \includegraphics[width=0.26\textwidth]{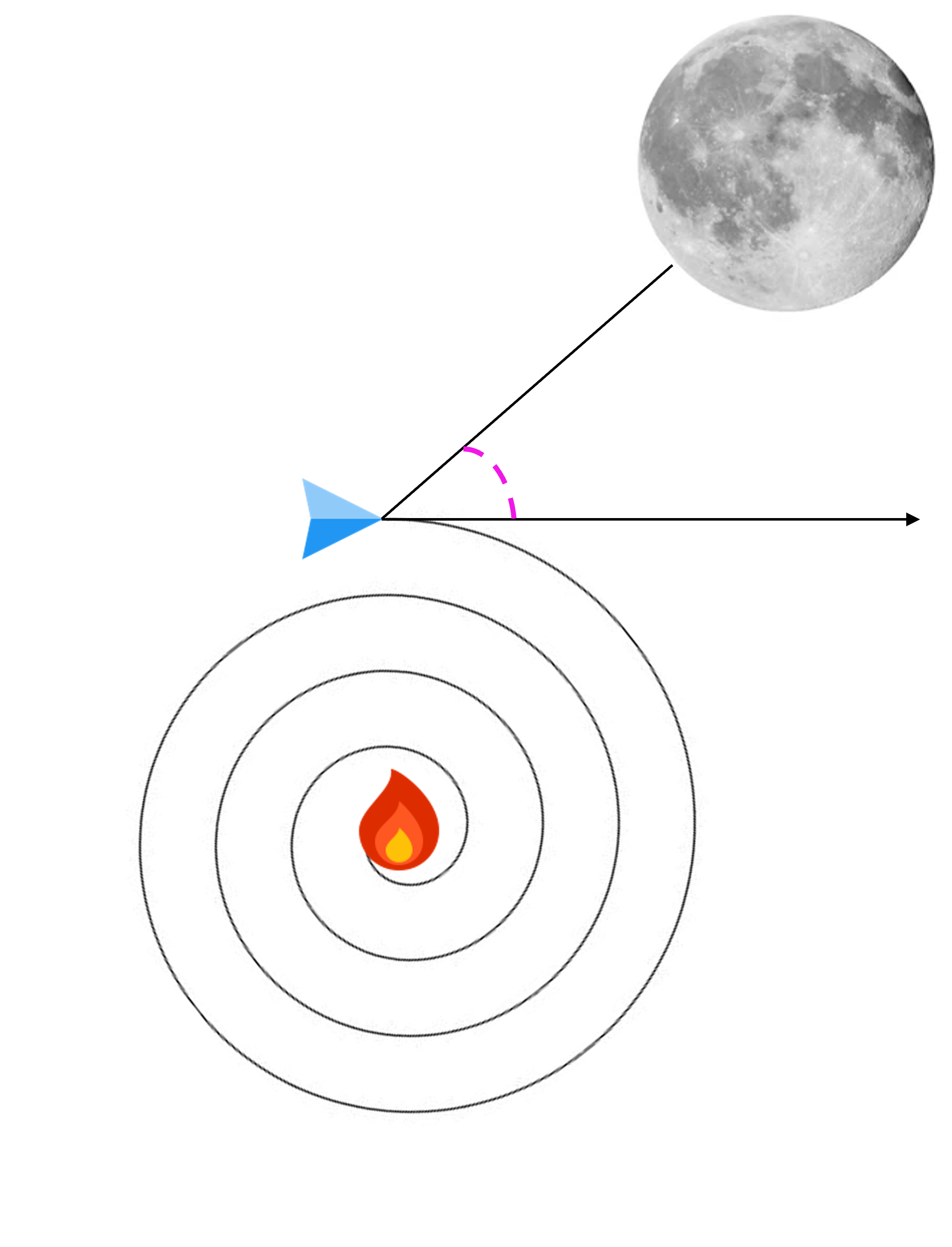}\\
    \caption{Flight mechanism of the moth. The upper portion illustrates the \textit{transverse orientation} mechanism. The lower portion illustrates the artificial light entrapment. The moth is represented using the blue arrow. }
    \label{fig:figure_transverse_orientation}
  \end{center}
\end{figure}

\textbf{Motivation and Rationale} As a nature-inspired optimiser, the Moth-flame Optimisation algorithm is population-based as it involves a population of moths. The moth has a special navigation mechanism called \textit{transverse orientation}, which they use as a flight path maintaining method. As shown in upper portion of the conceptual Figure \ref{fig:figure_transverse_orientation}, the moth attempts to maintain a fixed angle (marked in pink) between its flying direction and the moon, so that they can fly in a relatively straight path since the moon is far away from the moth. However, as illustrated in the lower part of Figure \ref{fig:figure_transverse_orientation}, the moth can sometimes confuse the artificial light with the moon. Then, it will try to maintain the transverse orientation mechanism with the light which is much closer than the moon, leading to the entrapment towards the light and eventually hit it. 

\vspace{0cm}

Inspired by this phenomenon, the Moth-flame Optimisation regards moths as search candidates, and treats the lights (flames) as potential optimal solutions. The Moth-flame Optimisation algorithm mimics the transverse orientation mechanism and hopes that the moth can reach the most optimal flame, which is regarded as the approximation to the global optima. By utilising a population of moths instead of a single one, the Moth-flame Optimiser possesses higher chance to avoid local optima entrapment and hence better approximates the global optima. 

\textbf{General Framework} \rw{The Moth-flame Optimisation algorithm works under the general framework of swarm-based algorithm \cite{genetic_algorithms_framework_mirjalili2019genetic}. The species population will firstly be initialised, then they will keep evolving, eliminating individuals with bad fitness and updating until the termination criteria are reached. Eventually, the fittest individual will survive and will be treated as the optimal solution. }

\textbf{Formulation} The Moth-flame Optimisation algorithm involves $n$ moths, each is a search candidate wandering in the search space. Each moth $M_n \in \mathbb{R}^d$ is a $d$ dimensional vector, where $d$ is the number of features to be optimised. Hence, it leads to the moth matrix $M$ with dimension $n \times d$, represented as follows: 
\begin{equation}\label{equ:moth_matrix}
  M = \begin{bmatrix}
    M_1\\
    M_2\\
    \vdots\\
    M_n
  \end{bmatrix}
    = \begin{bmatrix}
    m_{1, 1} & m_{1, 2} & \ldots & m_{1, d}\\
    m_{2, 1} & m_{2, 2} & \ldots & m_{2, d}\\
    \vdots   & \vdots   & \ddots & \vdots\\
    m_{n, 1} & m_{n, 2} & \ldots & m_{n, d}\\
    \end{bmatrix}
\end{equation}

\rw{A fitness function $f$ is required to evaluate the fitness of each moth  $M_n$ by taking it as input, and returns its fitness, i.e., the objective value. }The objective function has the following formulation: 
\begin{equation}\label{equ:objective_function_formulation}
  f: \mathbb{R}^d \rightarrow \mathbb{R}, f(M_n) = OM_n
\end{equation}
and hence, the corresponding fitness vector $OM$ is defined as follows: 
\begin{equation}\label{equ:moth_fitness_vector}
  OM = \begin{bmatrix}
    f(M_1)\\
    f(M_2)\\
    \vdots\\
    f(M_n)
  \end{bmatrix}
   = \begin{bmatrix}
    OM_1\\
    OM_2\\
    \vdots\\
    OM_n
  \end{bmatrix}
\end{equation}

\rw{In the Moth-flame Optimisation algorithm, the flames are not the real flames in the real world. Instead, they are set to be moths with top $k$ highest fitness values that have the right to survive (as in line $9$ in Algorithm \ref{algo:the_prodra_mf_optimisation_algorithm}), hence the flame matrix $F$ has dimension $k \times d$. }Without prior knowledge about which moth location is better, initially, the Moth-flame Optimisation algorithm randomly initialises the flame matrix $F$ with $k = n$. During iterations, the $k$ will be gradually decreased as the population evolves. The flame matrix $F$ is represented as follows: 
\begin{equation}\label{equ:the_flame_matrix}
  F = \begin{bmatrix}
    F_1\\
    F_2\\
    \vdots\\
    F_k
  \end{bmatrix}
    = \begin{bmatrix}
    f_{1, 1} & f_{1, 2} & \ldots & f_{1, d}\\
    f_{2, 1} & f_{2, 2} & \ldots & f_{2, d}\\
    \vdots   & \vdots   & \ddots & \vdots\\
    f_{k, 1} & f_{k, 2} & \ldots & f_{k, d}\\
    \end{bmatrix}
\end{equation}
and its corresponding fitness vector $OF$ is defined as follows:
\begin{equation}\label{equ:flame_fitness_vector}
  OF = \begin{bmatrix}
    f(F_1)\\
    f(F_2)\\
    \vdots\\
    f(F_k)
  \end{bmatrix}
   = \begin{bmatrix}
    OF_1\\
    OF_2\\
    \vdots\\
    OF_k
  \end{bmatrix}
\end{equation}
The details on how the Moth-flame Optimisation algorithm is integrated in the \emph{PECCO} model, i.e., what moth matrix $M$ stands for, etc., will be explained in Section \ref{sec:section_integration_of_the_moth_flame_optimiser}. 

\rw{As is previously mentioned, if there is no knowledge about which initial position is better, then a random initialisation will be applied} to generate both the moth matrix and the flame matrix using the following random generator: 
\begin{equation}\label{equ:random_initialiser}
  m_{i,j} = (ub(i) - lb(i)) * random() + lb(i)
\end{equation}
where $ub$ and $lb$ are the upper and lower bound of the range constraint, respectively. The $random()$ function is a random number generator with range in $[0, 1]$. 

\ul{Deficiency 1}: The Moth-flame Optimisation algorithm applies a random initialisation as it assumes there is no prior knowledge about which initial location is better. However, if prior knowledge presents, the random initialisation will degrade the performance. Besides, the random initialisation is not density-aware, i.e., the random initialisation may produce random vectors that are highly similar and hinder the diversity of the random population. An initial population with poor diversity will impair the benefit of population-based optimisers. 

\begin{figure}[!h]
  \begin{center}
    \includegraphics[width=0.3\textwidth]{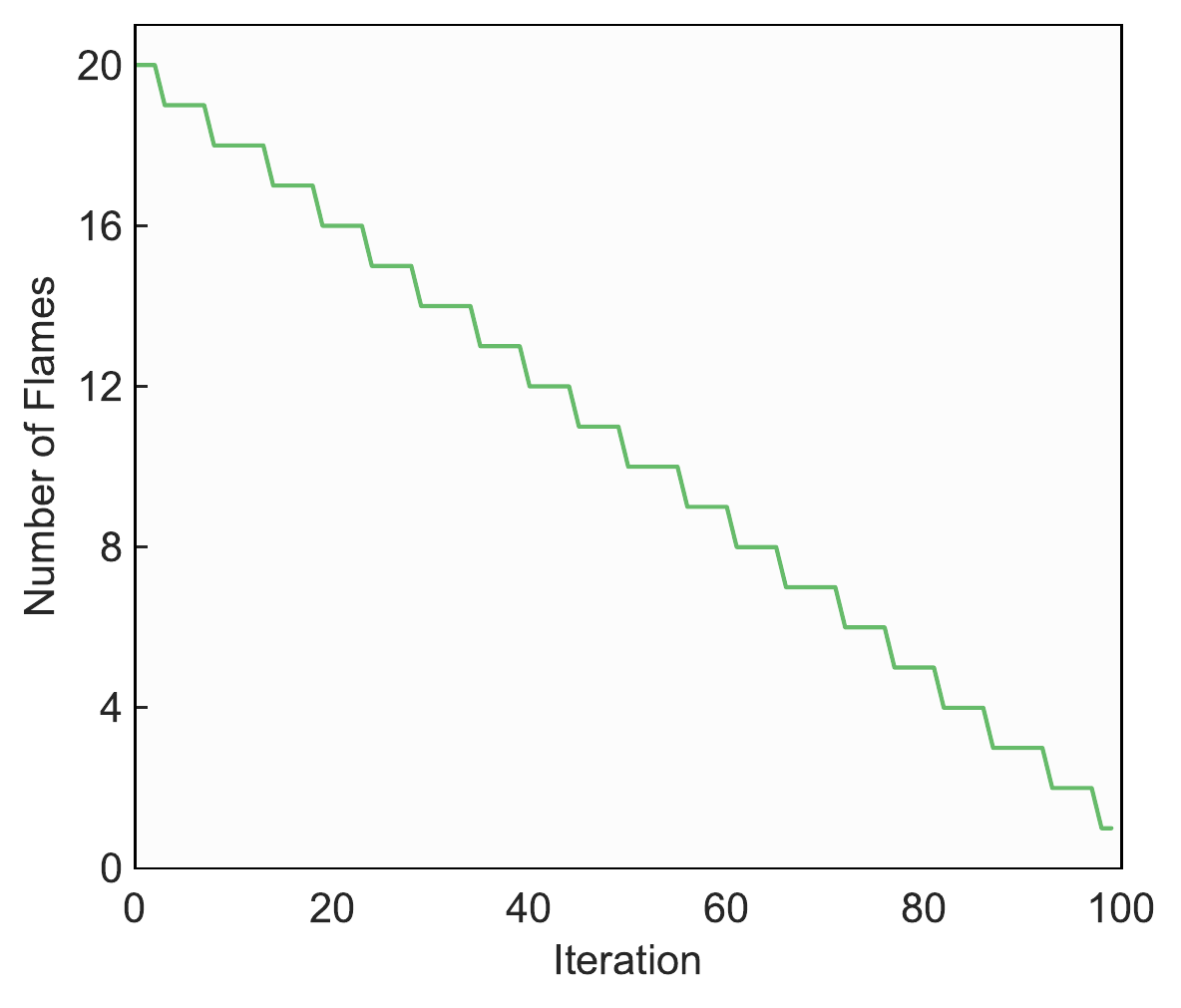}\\
    \caption{Illustration of the decreasing trend of the number of flames $k$, i.e., Equation \ref{equ:equation_step_round}. }
    \label{fig:figure_decreasing_k}
  \end{center}
\end{figure}

\begin{figure}[!h]
  \begin{center}
    \includegraphics[width=0.26\textwidth]{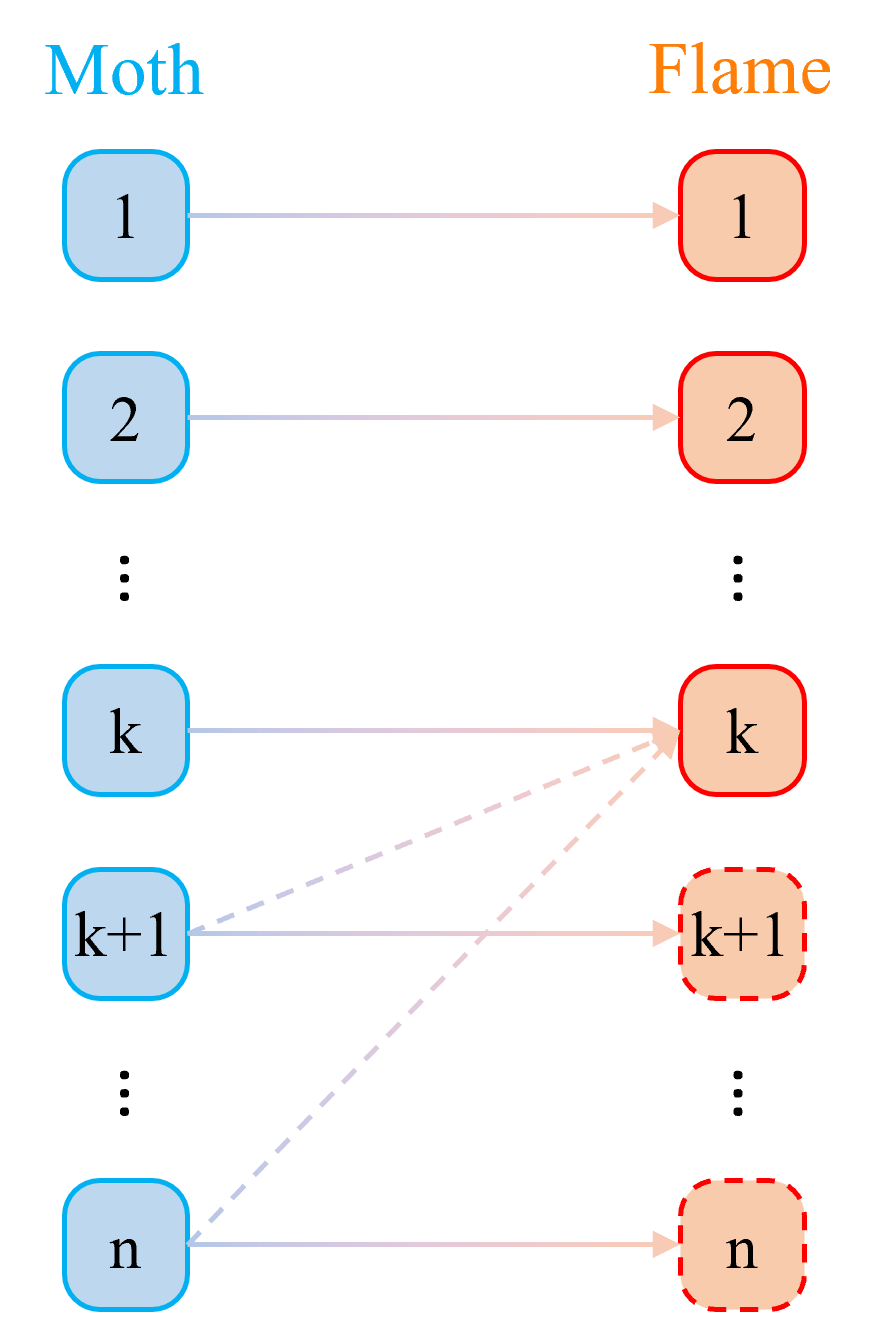}\\
    \caption{The original moth-flame pairing mechanism. The moths are represented using blue boxes while the flames are represented using orange boxes. }
    \label{fig:figure_moth_flame_pairing}
  \end{center}
\end{figure}

\textbf{Balancing Exploration and Exploitation} The Moth-flame Optimisation algorithm puts effort to balance between exploration and exploitation. Initially, there are $n$ moths and $n$ flames, each moth pursues its corresponding flame as illustrated by the solid arrows in Figure \ref{fig:figure_moth_flame_pairing}, which encourages exploration to avoid local optima stagnation as much as possible. During iterations, the moths will be sorted based on their fitness value in descending order, and the moths with top $k$ highest fitness value will survive while other moths will be eliminated as shown in Algorithm \ref{algo:the_prodra_mf_optimisation_algorithm}. The value of $k$ keeps decreasing based on the following formula during iterations so that the exploitation will be gradually emphasised: 
\begin{equation}\label{equ:equation_step_round}
  k = round(n - CI * \frac{n - 1}{MI})
\end{equation}
where $n$ is the initial number of moth/flame, $MI$ denotes the total number of iterations (Max iteration), $CI$ stands for the current iteration. Eventually, $k$ will decrease to $1$, the last survived flame is regarded as the optimal solution produced by the Moth-flame Optimisation algorithm. The decreasing trend of $k$ has been illustrated in Figure \ref{fig:figure_decreasing_k}. 

\vspace{0cm}

The number of flames $k$ keeps decreasing while the number of moths $n$ remains unchanged, the Moth-flame Optimisation algorithm therefore designs a moth-flame pairing mechanism as presented in Figure \ref{fig:figure_moth_flame_pairing} so that the moths can decide which target flame is designated for them to pursue. At the beginning, the number of moths and flames are equal, i.e., $n$, hence each moth will pursue its corresponding flame, i.e., $M_i \rightarrow F_i$, as represented by the solid arrows in Figure \ref{fig:figure_moth_flame_pairing}. During iterations, the value of $k$ will keep decreasing, hence the number of flames will be less than the number of moths. Under the MFO moth-pairing mechanism, the moth $M_i$ will still pursue its corresponding flame $F_i$ if flame $F_i$ still survives, or otherwise $M_i$ will chase the last flame $F_k$ as represented by the dashed arrows in Figure \ref{fig:figure_moth_flame_pairing}. 

\vspace{0cm}

\ul{Deficiency 2}: \rw{During moth evolvement, at any given time, moth $M_i$ will always pursue a single designated flame, which increases the chance of being trapped in the local optima. }

\ul{Deficiency 3}: \rw{During moth evolvement, those moths that have their corresponding flame being eliminated will always pursue the last surviving flame, which is the flame with the worst fitness. Neither pursuing the worst flame nor letting lots of moths pursuing a single flame is a reasonable design. }

\begin{figure}[!h]
  \begin{center}
    \includegraphics[width=0.3\textwidth]{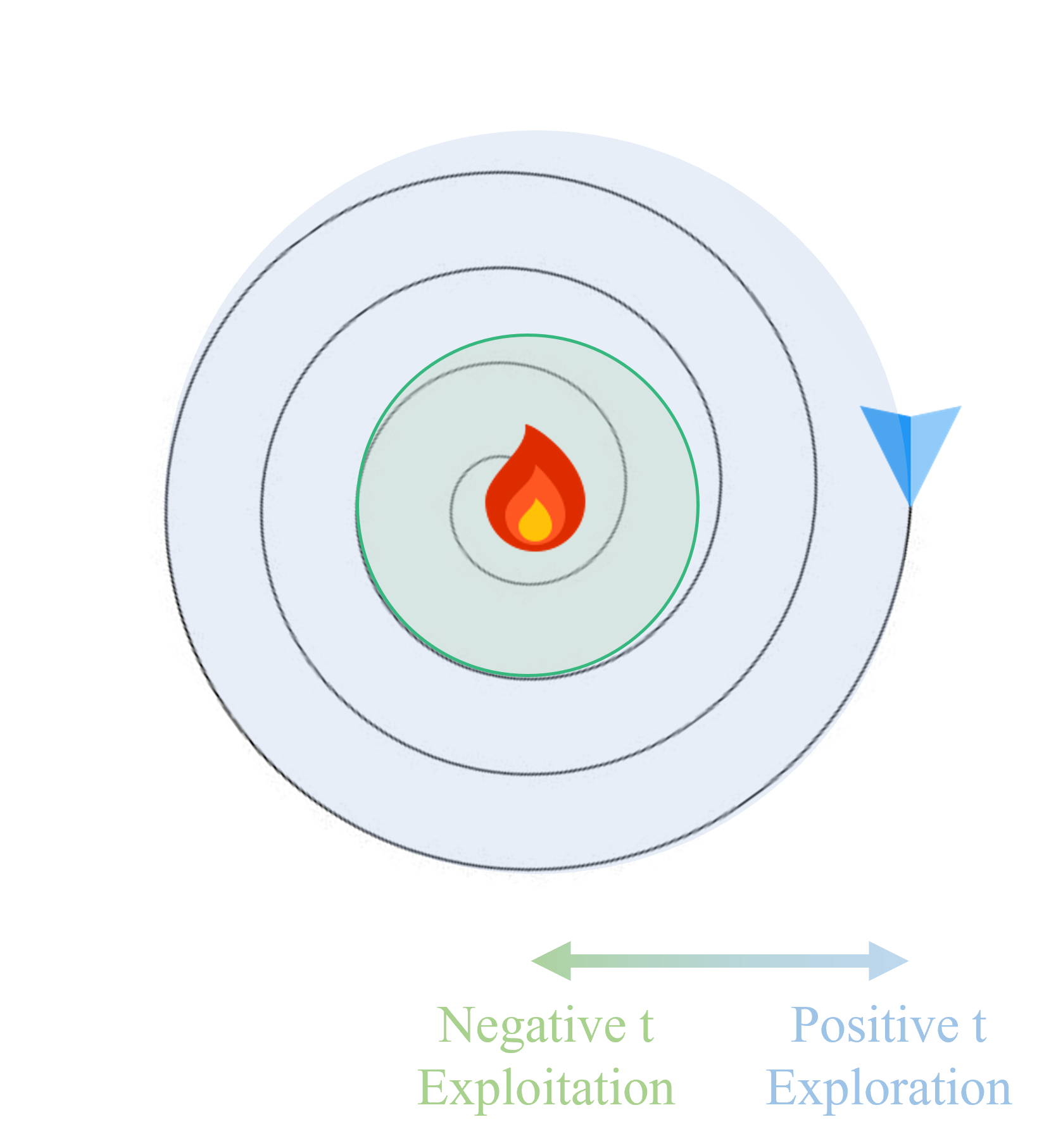}\\
    \caption{Illustration of the exploration vs exploitation of the Moth-flame Optimisation algorithm. }
    \label{fig:figure_exploration_exploitation}
  \end{center}
\end{figure}

\rw{In terms of the moth updating mechanism, the Moth-flame Optimisation algorithm mimics the transverse orientation based on the following equation: }
\begin{equation}\label{equ:equation_spiral_update_equation}
  U(M_i, F_j) = D_{i,j} \times e^{bt} \times cos(2 \pi t) + F_j
\end{equation}
where $F_j$ is the paired flame designated for $M_i$ to pursue, $t$ is a random number in range $[r, 1]$, $r$ is a random number that will linearly decrease from $-1$ to $-2$ during iterations, $b$ is the shape parameter, and $D_{i, j}$ denotes the $L1$-distance between $M_i$ and $F_j$ which is defined as follows: 
\begin{equation}\label{equ:equation_distance_calculation}
  D_{i,j} = |F_j - M_i|
\end{equation}
The shape of an example updating path, i.e., the spiral shape, has been illustrated in Figure \ref{fig:figure_transverse_orientation} and \ref{fig:figure_exploration_exploitation}. 

\vspace{0cm}

Specifically, the $t$ parameter decides how close to the flame will the moth's terminal position be. As illustrated in Figure \ref{fig:figure_exploration_exploitation}, a $t$ value that is closer to $1$ will let the moth ends up with a position that is farther from the flame (the blue shaded area), which emphasises exploration. On the other hand, a negative $t$ value will draw the moth closer towards its target flame as indicated by the green shaded area in Figure \ref{fig:figure_exploration_exploitation}, which encourages exploitation. Since t is within the range of $[r, 1]$, initially, $r$ has value $-1$ which promotes exploration by avoiding the moth to be too close to the flame. As the process evolves, $r$ linearly decreases from $-1$ to $-2$, which gradually encourages exploitation over exploration. 

\textbf{Termination} \rw{The termination criterion is when there is only one flame remaining. It will be treated as the optimal solution. }

\textbf{Advantage and Applicability} In summary, the Moth-flame Optimisation algorithm has the following advantages which make it applicable in our case: 
\begin{itemize}
  \item Since the \emph{PECCO} model is hard in nature, applying this population-based algorithm with multiple search candidates while enabling the balance between exploration and exploitation will possess higher chance to approximate the global optima. 
  \item Since the objective function in the \emph{PECCO} model is not differentiable, the Moth-flame Optimiser becomes applicable as it evaluates each search candidate using the fitness function and therefore is gradient-free. 
  \item \rw{Compared with its counterparts, the Moth-flame Optimiser achieves superior optimisation results and converges in an efficient manner. }
\end{itemize}

\textbf{Room for Improvements} As is aforementioned, the Moth-flame Optimisation algorithm suffers from three deficiencies, which leave us with room for improvements. We propose three new mechanisms to tackle these deficiencies as follows: 
\begin{itemize}
  \item The profit, cost and density-aware initialiser $\rightarrow$ Deficiency 1
  \item The  dynamic hierarchical flaming mechanism $\rightarrow$ Deficiency 2
  \item The lifetime-enabled moth-flame pairing strategy $\rightarrow$ Deficiency 3
\end{itemize}
Together, these mechanisms form the improved \emph{PECCO-MFI} algorithm. The details will be presented in Section \ref{sec:section_the_improvements}.

%%%%%%%%%%%%%%%%%%%%%%%%
%%% Model and Method %%%
%%%%%%%%%%%%%%%%%%%%%%%%

\section{Model and Method}
\label{sec:section_model_and_method}

\rw{In this section, the problem formulation will be provided, followed by the presentation of the proposed \emph{PECCO} optimisation model, in which the profit and cost component of the \emph{PECCO} model will be explained. }Then, we will explain how the Moth-flame Optimisation algorithm is improved and integrated to form our \emph{PECCO-MFI} algorithm. 

\subsection{Problem Formulation}
\label{sec:section_problem_formulation}

\rw{In the edge-cloud environment, there are cloud nodes and edge devices (nodes), }with a connection topology to form a connected graph. Hence, we formulate the problem as a graph $G = (V, E)$ where $V$ stands for a set of cloud/edge nodes and $E$ represents a set of communication links. There are in total $N$ computing nodes, in which it contains $I$ cloud nodes and $J$ edge nodes, hence we have 
\begin{equation}\label{equ:equation_v_definition}
  \begin{split}
    & V^C = \{V_1^C, V_2^C, \cdots, V_I^C\}, V^E = \{V_{I+1}^E, V_{I+2}^E, \cdots, V_{I+J}^E\}\\
    & V = V^C \cup V^E, V^C \cap V^E = \emptyset, N = I + J
  \end{split}
\end{equation}
\rw{Note that we can simplify the notation $V^X_n$, $X \in \{C, E\}$ to be $V_n$ as the range of subscript $n$ can tell whether the node belongs to the cloud or the edge. }

For each computing node $V_n$, it has the following properties. Firstly, each computing node is capable of handling certain capacity of computation tasks. Hence $Cap_{V_n}\_max$ denotes the maximum number of units of computation workload that node $V_n$ is capable of handling, while $Cap_{V_n}\_min$ stands for the minimum workload of node $V_n$ when it is idle. We assume that no node can be overloaded by computation tasks. By considering the capacity of each node, it can also indirectly model other performance factors such as power consumptions. 

As for edges $E$ in the graph $G$, there are in total $Q$ edges, denoted as $E_q$, or interchangeably $E_{<V_s,V_t>}$, which stands for edge $E_q$ is an edge that starts from node $V_s$ and points to node $V_t$. To make the heterogeneous model more generalisable, the length of each edge is also considered instead of being ignored as in \cite{heto_du2020algorithmics,larac_juttner2001lagrange}, and is denotes as $L_{E_q}$ (or $L_{E_{<V_s,V_t>}}$ using the interchangeable notation). 

In terms of tasks to be executed, there are in total $K$ of them, each task $T_k$ has a property $WL_{T_k}$ that represents how many units of computation workload does task $T_k$ have. Each task can only be allocated on either a cloud node or an edge node, and a task is allowed to stay if its initial allocation is good enough. Hence, we define $\mathcal{A}_{T_k}^I$ and $\mathcal{A}_{T_k}^O$ to be the initial and offloaded allocation of task $T_k$, which satisfies 
\begin{equation}\label{equ:equation_ai_ao}
  \mathcal{A}_{T_k}^I, \mathcal{A}_{T_k}^O \in V^X, X \in \{C, E\}
\end{equation}
Moreover, vector $\mathcal{A}^O$ is defined to represent the offloaded allocations of all $K$ tasks as follows: 
\begin{equation}\label{equ:equation_A_O}
  \mathcal{A}^O = \begin{bmatrix}
    \mathcal{A}^O_{T_1}\\
    \mathcal{A}^O_{T_2}\\
    \vdots\\
    \mathcal{A}^O_{T_K}
  \end{bmatrix}
\end{equation}
Besides, the \emph{PECCO} model applies a heterogeneous cost model for tasks. Instead of applying a homogeneous task cost as in \cite{heto_du2020algorithmics}, for each task, it has different costs $C_{T_k}^C$ if it is executed on the cloud, or $C_{T_k}^E$ if being allocated to the edge. \rw{By utilising a heterogeneous cost model for each task, it can reflect that different tasks can have different costs when being allocated to different sides, which makes the model more realistic. }To make the model profit-oriented, each task is also associated with two profits, i.e., $P_{T_k}^C$ and $P_{T_k}^E$, \rw{which stand for the profit gained of completing task $T_k$ on the cloud and edge, respectively. }By jointly considering profit and cost, it makes the proposed \emph{PECCO} model become profit and cost-oriented. 

\subsection{\emph{PECCO} Cost Model}
\label{sec:section_the_pecco_edge_cloud_computation_offloading_cost_model}

The \emph{PECCO} model is a multi-factored model that jointly considers both the generalised heterogeneous communication cost and the heterogeneous computation cost.  

\subsubsection{Generalised Heterogeneous Communication Cost Model}
\label{sec:section_generalised_heterogeneous_communication_cost_model}

Considering that previously proposed communication cost models in past researches suffered from several drawbacks (e.g., applied the unrealistic symmetric and cost-free assumption, failed to consider communication distance, \rw{applied homogeneous communication costs between node pairs, }etc. ) which made them become hardly generalisable in practice, it naturally leads to the rationale of our generalised heterogeneous communication cost model. 

There are in general four types of communication costs, i.e., $w^{CC}$, $w^{EE}$, $w^{CE}$ and $w^{EC}$. The $CE$ here for instance represents the communication from a cloud node to an edge node. Inside each type of communication cost, it can also have different costs between different nodes, which is denotes as $w^{XX}_{<V_s,V_t>}$. For example, $w^{EC}_{<V^E_j,V^C_i>}$ denotes the communication cost from edge node $V^E_j$ to the cloud node $V^C_i$. \rw{As different nodes may work under different conditions like being operated by different service providers, }the communication cost between node pairs can be different even if they are situated on the same side. \rw{Therefore, the model is more realistic in practice. }This kind of generalisation also offers convenience to represent communication failures for instance, by setting the edge-wise communication cost to be a large value. \rw{The communication cost function $C_{E_{<V_s,V_t>}}$ is formulated as follows: }
\begin{equation}\label{equ:equation_communication_indicator_function}
  \begin{split}
    & C_{E_{<V_s,V_t>}} = sum(L_{E_{<V_s,V_t>}}
    \times [1_{V_s, V_t \in V^C}, 1_{V_s, V_t \in V^E}, 1_{V_s \in V^C, V_t \in V^E}, 1_{V_s \in V^E, V_t \in V^C}] 
    \odot [w^{CC}_{<V_s,V_t>}, w^{EE}_{<V_s,V_t>}, w^{CE}_{<V_s,V_t>}, w^{EC}_{<V_s,V_t>}])
  \end{split}
\end{equation}
inside it, for instance, $1_{V_s \in V^C, V_t \in V^E}$ will return 1 if $V_s$ is a cloud node and $V_t$ is an edge node, and will return 0 otherwise, other indicator functions carry the similar meaning. The $\odot$ represents the element-wise multiplication operator and $\times$ is the scalar multiplication operator. Hence, the communication cost function $C_{E_{<V_s,V_t>}}$ will return the length of the inputted edge times the corresponding cost of that type of communication so that the heterogeneous communication costs between node pairs can be considered. 

After defining the generalised heterogeneous communication cost model, the optimal cost path between any pairs of computing nodes can be pre-computed using the shortest path algorithm. The optimal cost path between node $V_i$ and $V_j$ is denoted as $OCP_{<V_i,V_j>}$, and therefore we can define the optimal communication cost from node $V_i$ to node $V_j$ as follows: 
\begin{equation}\label{equ:generalised_node_wise_communication_cost}
  COMM_{<V_i,V_j>} = \sum_{E_{<V_s,V_t>} \in OCP_{<V_i,V_j>}} C_{E_{<V_s,V_t>}}
\end{equation}
and therefore, the total communication cost is defined as follows:
\rw{\begin{equation}\label{equ:generalised_total_communication_cost}
  COMM(G, T, \mathcal{A}^I, \mathcal{A}^O) = \argmin_{\mathcal{A}^O}\{\sum_{k=1}^K COMM_{<\mathcal{A}^I_{T_k}, \mathcal{A}^O_{T_k}>}\}
\end{equation}}

The optimisation algorithm should find an optimal offloading strategy $\mathcal{A}^O$ to offload task $T_k$ so that it can achieve a communication cost $COMM(G, T, \mathcal{A}^I)$ as low as possible. \rw{In summary, the proposed generalised heterogeneous communication cost model overcomes the drawbacks of previously proposed communication models and has the following benefits: }

\begin{itemize}
  \item It no longer ignores the communication cost between nodes on the same side (i.e., cloud-cloud, edge-edge). 
  \item The asymmetry between communication costs is considered, cost from cloud to edge and from edge to cloud communication can be heterogeneous. 
  \item It considers distances when modelling communication cost between two nodes. 
  \item It allows different node pairs to have heterogeneous communication costs. 
\end{itemize}

\subsubsection{Heterogeneous Computation Cost Model}
\label{sec:section_generalised_computation_cost_model}

\rw{Next, the heterogeneous computation cost model is defined which also takes the heterogeneities between computing tasks into account. }Generally, for each task $T_k$, it possesses cost $C_{T_k}^C$ and $C_{T_k}^E$, which is the cost of executing task $T_k$ on the cloud and edge, respectively. Due to the diversity of tasks, $C_{T_k}^C$ is not necessarily lower than $C_{T_k}^E$. The previously proposed homogeneous computation cost model for tasks is infeasible, it is unreasonable to assume that all tasks share exactly the same computation cost when being executed on a single side. Hence, the computation cost model for tasks we considered in the \emph{PECCO} is more generalisable. The overall computation cost is formulated as follows:
\rw{\begin{equation}\label{equ:computation_cost}
  \begin{split}
    & COMP(G, T, \mathcal{A}^O) = 
    \argmin_{\mathcal{A}^O} \{\sum^K_{k=1} (1_{\mathcal{A}^O_{T_k} \in V^C} \times C^C_{T_k} + 1_{\mathcal{A}^O_{T_k} \in V^E} \times C^E_{T_k})\}
  \end{split}
\end{equation}}
where the indicator function $1_{\mathcal{A}^O_{T_k} \in V^C}$ will return 1 if the allocation for task $T_k$ $\mathcal{A}^O_{T_k}$ is a cloud node, and will return 0 otherwise, similar for $1_{\mathcal{A}^O_{T_k} \in V^E}$. 

\subsection{\emph{PECCO} Profit Model}
\label{sec:section_pecco_profit_model}

Different from previously proposed works, the proposed \emph{PECCO} model is not only cost-oriented, but also profit-oriented. \rw{For each task $T_k$, it has profit $P_{T_k}^C$ and $P_{T_k}^E$ when being executed on the cloud and edge, respectively. The overall profit is formulated as follows: }
\rw{\begin{equation}\label{equ:profit}
  \begin{split}
    & PROFIT(G, T, \mathcal{A}^O) = 
    \argmin_{\mathcal{A}^O} \{\sum^K_{k=1} (1_{\mathcal{A}^O_{T_k} \in V^C} \times P^C_{T_k} + 1_{\mathcal{A}^O_{T_k} \in V^E} \times P^E_{T_k})\}
  \end{split}
\end{equation}}

\subsection{Overall Optimisation Objective}
\label{sec:section_overall_objective_function}

\rw{Finally, the \emph{PECCO} optimisation model will integrate the aforementioned cost and profit model to become profit and cost-oriented. The objective function is defined as follows: }
\rw{\begin{equation}\label{equ:overall_objective}
  \begin{split}
    & Obj(G, T, \mathcal{A}^I) = 
    \argmin_{\mathcal{A}^O} \{(COMM(G, T, \mathcal{A}^I) + COMP(G, T, \mathcal{A}^O))
    + \lambda \times PROFIT(G, T, \mathcal{A}^O)\}
  \end{split}
\end{equation}}
\rw{where $\lambda$ is a ratio parameter being set to a negative value to integrate the profit into the objective to be minimised. }By having $\lambda$, the objective function can minimise the cost and simultaneously maximise the profit. 

By jointly optimising the profit and cost-oriented optimisation model \emph{PECCO}, we can find a solution that can jointly optimise costs and the profit as much as possible. 

\subsection{The \emph{PECCO-MFI} Optimiser}
\label{sec:section_the_moth_flame_optimiser}

In this section, we will introduce the proposed improved Moth-flame Optimiser with detailed explanations to the improvements we made. Then, how the improved Moth-flame Optimisation algorithm is integrated to optimise the \emph{PECCO} model is explained, i.e., what moths stand for in the \emph{PECCO-MFI} algorithm, how are tasks offloaded based on the optimised allocation strategy $\mathcal{A}^O$, etc. 

\subsubsection{Algorithm Improvement}
\label{sec:section_the_improvements}

\rw{To tackle the deficiencies mentioned in Section \ref{sec:section_the_moth_flame_optimisation_algorithm} and therefore boost the performance, }we propose an improved Moth-flame Optimiser called \emph{PECCO-MFI} with three improvements to tackle three deficiencies, respectively. 

\begin{algorithm}[!ht]
    \begin{algorithmic}[1]
        \Require
            \StateX Number of search candidates (moths) $nsa$, 
            \StateX Edge-cloud graph $G$, 
            \StateX Tasks $T_k \in T$, 
            \StateX Allocation upper bound $ub$, 
            \StateX Objective function $Obj()$ as defined in Equation \ref{equ:overall_objective}, 
            \StateX Initial allocation $\mathcal{A}^I$
        \Ensure Profit, cost and density-aware moth initialisation with dimension $nsa \times N$
        \For{$T_k$ \textup{in} $T$}
            \State Calculate costs based on Equation \ref{equ:generalised_total_communication_cost} and \ref{equ:computation_cost}
            \State Calculate the profit based on Equation \ref{equ:profit}
            \State Calculate the profit and cost-oriented objective based on Equation \ref{equ:overall_objective}
        \EndFor
        \For{$i$ \textup{in} $range(nsa \times 1.5)$}
            \For{$k$ \textup{in} $range(K)$}
                \If{allocate task $T_k$ to the cloud side yield a lower objective value}
                    \State Store random number in range $[0, \frac{ub}{2})$ into $\mathcal{A}^I_i$
                \Else
                    \State Store random number in range $[\frac{ub}{2}, ub]$ into $\mathcal{A}^I_i$
                \EndIf
            \EndFor
        \EndFor
        \While{$len(\mathcal{A}^I) \neq nsa$}
            \State Find pair $(\mathcal{A}^I_i, \mathcal{A}^I_j)$ with minimum intra-pair L$2$ distance
            \State Add $\frac{\mathcal{A}^I_i + \mathcal{A}^I_j}{2}$ into $\mathcal{A}^I$
            \State Remove both $\mathcal{A}^I_i$ and $\mathcal{A}^I_j$ from $\mathcal{A}^I$
        \EndWhile
        \State \Return $\mathcal{A^I}$
    \end{algorithmic}
\caption{The profit, cost and density-aware moth initialiser $initialiser(nsa, G, T, ub, obj, \mathcal{A}^I)$ of the \emph{PECCO-MFI} Algorithm}
\label{algo:the_profit_aware_initialisation}
\end{algorithm}

\textbf{Improvement 1 (Profit, Cost and Density-aware Moth Initialiser)}: To tackle the \ul{Deficiency 1: trivial random initialisation}, we design a new moth initialiser that is profit, cost and density-aware, as shown in Algorithm \ref{algo:the_profit_aware_initialisation}. 

\ul{Profit and Cost-awareness}: The newly designed moth initialiser will allocate tasks (elements in each moth) to cloud or edge side based on their profit and cost. It is natural to allocate services to the side in which they possess a lower profit-cost objective than the other side. \rw{Conversely, if the allocation is done in the reversed way, then this task is likely to be migrated during the optimisation process, which therefore incurs unnecessary costs. }The profit and cost-aware moth initialisation mechanism has been shown in line $7$ - $13$ in Algorithm \ref{algo:the_profit_aware_initialisation}. By utilising this prior knowledge, the improved Moth-flame Optimisation algorithm is reasonable to outperform its knowledgeless random counterpart. 

\ul{Density-awareness}: The rationale of applying the population-based paradigm is to maximise the chance of approximating the global optima as close as possible. However, if some moths are initialised to have a close proximity, the benefit of the population-based paradigm will be greatly hindered. \rw{To ensure the initialised moths  have rich diverity, }the newly designed initialiser will generate more moth vectors than required, then it will iteratively remove the closest pair of moths and keep the average of them. The procedure has been given in line $15$ - $19$ in Algorithm \ref{algo:the_profit_aware_initialisation}. The removal will be continued until the number of moth vectors is satisfied as required. By leveraging this mechanism, moth vectors that are initialised to be too close will be merged into a new one to prevent the performance degradation from happening. Therefore, the proposed moth initialiser is density-aware. 

\begin{algorithm}[!ht]
    \begin{algorithmic}[1]
        \Require 
            \StateX Current iteration $CI$, 
            \StateX Max iteration $MI$
        \State Define $F_\mathbb{1}$, $F_\mathbb{2}$, $F_\mathbb{3}$ as the flames with top $3$ fitness values, respectively
        \State $\omega \leftarrow \frac{CI}{MI}$
        \If{moth $M_i$'s corresponding flame $F_i$ still survives}
            \If{there are $\geq 3$ flames survive}
                \State $M_i$ will pursue $\frac{F_i + \omega \times F_\mathbb{1} + \omega \times F_\mathbb{2} + \omega \times F_\mathbb{3}}{1 + 3 \times \omega}$
            \ElsIf{there are $2$ flames survive}
                \State $M_i$ will pursue $\frac{F_i + \omega \times F_\mathbb{1} + \omega \times F_\mathbb{2}}{1 + 2 \times \omega}$
            \EndIf
        \Else
            \State $\tau \leftarrow random(0, 1)$, where $\tau$ is the lifetime parameter
            \If{$\tau$ > $0.8$}
                \State $F_{\mathcal{I}} \leftarrow$ a randomly initialised flame
            \Else
                \State $\mathcal{I} \leftarrow random(0, k)$
            \EndIf
            \If{there are $\geq 3$ flames survive}
                \State $M_i$ will pursue $\frac{F_{\mathcal{I}} + \omega \times F_\mathbb{1} + \omega \times F_\mathbb{2} + \omega \times F_\mathbb{3}}{1 + 3 \times \omega}$
            \ElsIf{there are $2$ flames survive}
                \State $M_i$ will pursue $\frac{F_{\mathcal{I}} + \omega \times F_\mathbb{1} + \omega \times F_\mathbb{2}}{1 + 2 \times \omega}$
            \EndIf
        \EndIf
    \end{algorithmic}
\caption{\rw{The dynamic hierarchical flaming mechanism and the lifetime-enabled moth-flame pairer $enhanced\_pairer(CI, MI)$ of the \emph{PECCO-MFI} Algorithm}}
\label{algo:the_enhanced_moth_flame_pairing_mechanism}
\end{algorithm}

\textbf{Improvement 2 (Dynamic Hierarchical Flaming Mechanism)}: To deal with the \ul{Deficiency 2: Single moth-flame pairing}, a dynamic hierarchical flaming mechanism is applied. \rw{As pursuing a single flame will lead to higher risk of being trapped in local optima, inspired by the social hierarchy possessed in the moth species, the moths with top $3$ highest fitness values will be regarded as leaders, }which will provide guiding reference for other moths to pursue. Hence, instead of pursuing a single flame, in the newly designed algorithm, each moth will chase the linear combination of its designated flame and the leader flames, as shown in line $3$ - $8$ in Algorithm \ref{algo:the_enhanced_moth_flame_pairing_mechanism}. 

\ul{Exploration and Exploitation Balance}: At the beginning of the training process, putting too much dependency on the top $3$ flames will incur risk of local optima stagnation as the performance of flames at the beginning is not promising enough. Hence, an adjusting factor $\omega$ is introduced which linearly increases from $0$ to $1$ during the training process as indicated in line $2$ in Algorithm \ref{algo:the_enhanced_moth_flame_pairing_mechanism}. \rw{After applying the adjusting factor $\omega$ as in line $5$ and $7$ in Algorithm \ref{algo:the_enhanced_moth_flame_pairing_mechanism}, exploration will be encouraged at the beginning by putting less emphasise on the top $3$ flames since initially the value of $\omega$ is small. }As the training progresses, $\omega$ will gradually increase, which will emphasise more exploitation, since the guiding reference of top $3$ flames will be gradually reinforced as the $\omega$ keeps growing. As such, the utilisation of adjusting factor $\omega$ in the newly designed dynamic hierarchical flaming mechanism will balance between exploration and exploitation. 

\textbf{Improvement 3 (Lifetime-enabled Moth-Flame Pairing Strategy)}: Finally, to solve the \ul{Deficiency 3: naive moth-flame pairing}, we design a fairer pairing strategy as indicated in line $9$ - $21$ in Algorithm \ref{algo:the_enhanced_moth_flame_pairing_mechanism}. Instead of letting all moths whose corresponding flames are eliminated to chase the last surviving flame, we introduce a lifetime parameter $\tau$ with lifetime threshold set as $0.8$. Due to the elimination of their unpromising flames, these moths are not promising themselves and hence the lifetime parameter $\tau$ is used to decide whether certain moth will be re-initialised, i.e., starting a new lifetime. Hence, as indicated in line $11$ - $12$ in algorithm \ref{algo:the_enhanced_moth_flame_pairing_mechanism}, if the randomly generated lifetime parameter $\tau$ is higher than the lifetime threshold, the moth will start a new lifetime by pairing with a newly initialised flame. Otherwise, the moth will continue its lifetime, and the algorithm will let it to pair with a randomly selected survived flame to promote a fairer exploration. By randomly pairing with a survived flame, these moths will fairly explore all possible survived flames instead of all exploiting the worst-fitted flame. During the re-pairing process, the aforementioned dynamic hierarchical flaming mechanism will be utilised again to provide better exploration and exploitation balancing while enabling the guiding reference of top $3$ flames as in line $16$ - $20$ in Algorithm \ref{algo:the_enhanced_moth_flame_pairing_mechanism}. By utilising this enhanced moth-flame pairing strategy, the exploration of the algorithm will be further encouraged and hence leading to a higher chance to approach the global optima. 

\subsubsection{Integration of the Moth-flame Optimiser}
\label{sec:section_integration_of_the_moth_flame_optimiser}

\begin{algorithm}[!ht]
    \begin{algorithmic}[1]
        \Require
            \StateX Shape parameter $b$, 
            \StateX Number of search candidates (moths) $nsa$, 
            \StateX Allocation upper bound $ub$, 
            \StateX Objective function $Obj()$ as defined in Equation \ref{equ:overall_objective}
        \Ensure Resource allocation strategy $\mathcal{A}^O$ of computation tasks, which is the best flame $F$
        \State Initialise moth matrix $M \leftarrow initialiser(nsa, G, T, ub, obj, \mathcal{A}^I)$ (in Algorithm \ref{algo:the_profit_aware_initialisation})
        \State $OM \leftarrow Obj(M)$
        \While{$len(F) \neq 1$}
            \State Update $k$
            \State $OM \leftarrow obj(M)$
            \If{$CI$ = 1}
                \State $F \leftarrow M.sortBy(OM)$
            \Else
                \State $F \leftarrow M.sortBy(OM)[0:k]$
            \EndIf
            \State Update moth-flame pairing using $enhanced\_pairer(CI, MI)$ (Algorithm \ref{algo:the_enhanced_moth_flame_pairing_mechanism})
            \For{$i$ \textup{in} $range(nsa)$}
                \For{$j$ \textup{in} $range(K)$}
                    \State Update $r$ and $t$
                    \State Calculate $D$ with respect to the paired moth and flame using Equation (\ref{equ:equation_distance_calculation})
                    \State Update moth position using Equation (\ref{equ:equation_spiral_update_equation})
                \EndFor
            \EndFor
        \EndWhile
        \State $OF \leftarrow obj(F)$
        \State \Return $F$, $OF$
    \end{algorithmic}
\caption{Workflow of the \emph{PECCO-MFI} Algorithm}
\label{algo:the_prodra_mf_optimisation_algorithm}
\end{algorithm}

\rw{We apply the improved Moth-flame Optimiser \emph{PECCO-MFI} to optimise the \emph{PECCO} model. The pseudocode of the algorithm has been presented in Algorithm \ref{algo:the_prodra_mf_optimisation_algorithm}. }In the \emph{PECCO-MFI} algorithm, each moth vector $M_i$ is a $1 \times K$ vector, where $K$ is the number of tasks waiting to be allocated. The values in the moth vector $M_i$ are within the range of $[0, ub]$, where $ub$ is a constant. If the value is in range $[0, \frac{ub}{2})$, it indicates that this task will be executed to the cloud side, otherwise, this task will be offloaded to the edge side. Then, the algorithm will find the computing node with the cheapest communication cost in the designated side and allocate the task to that computing node. If the computing node will be overloaded by taking this task, the algorithm will find the node at the designated side with the second cheapest communication cost and so on. \rw{Eventually, the task will either be allocated to a computing node without causing overloading, or it will not be satisfied due to workload unavailability. }

%%%%%%%%%%%%%%%%%%
%%% Experiment %%%
%%%%%%%%%%%%%%%%%%

\section{Experiment}
\label{sec:section_experiment}

In this section, we will introduce our experimental setup and the dataset we utilised during experiments. Then, experimental results will be presented and explained to testify to the superiority of the proposed method. Specifically, objective values yielded by different methods are compared, followed by the comparison of profit, cost and profit-cost ratio to demonstrate the effectiveness of the \emph{PECCO-MFI} algorithm in terms of joint profit and cost optimisation. Finally, to demonstrate the \emph{PECCO-MFI} algorithm offloads computation tasks wisely, the task allocation and the resource utilisation are compared and analysed. 

\subsection{Dataset, Parameter and Experimental Setup}
\label{sec:section_experimental_setup_and_dataset}

\textbf{Dataset}. The dataset we use to simulate the edge-cloud environment \cite{huang2019dp_greedy} is the Sydney train station parking dataset obtained from the Open Data Portal provided by the New South Wales Government Department of Transportation \cite{nsw_parking}. The dataset contains the parking lot availability information at each train station in Sydney, Australia. Train stations in City of Sydney are treated as cloud nodes, and suburban train stations act as edge nodes. Each station has a parking with limited available parking lots, which represents the capacity of the node. The length of connected edges are the length of the shortest road between train stations. The dataset contains communication heterogeneity as paths between different nodes can have different charges due to factors such as toll roads, etc. Parking requests are simulated as tasks which will be allocated to nodes. We pick $20$ stations from the City of Sydney to act as cloud nodes, and $30$ stations from Sydney suburban areas as edge nodes, numbered from $1$ to $20$, and from $21$ to $50$, respectively. We utilise $200$ tasks, each with certain amount of parking requests that will be treated as workloads. Parking in the city or in the suburban area can yield different parking fares, which serves as the cost of the task when being executed on the cloud side and edge side, respectively. Finally, successfully allocating each parking task will earn certain profit. 

\textbf{Parameter Setting}. We now introduce parameter settings in two parts: \emph{PECCO} optimisation model parameter settings, as well as improved Moth-flame Optimiser parameter settings. 

\ul{\textit{PECCO Optimisation Model Parameter Setting}}. To reflect the communication heterogeneity, we set the communication cost $w^{CC}$, $w^{CE}$, $w^{EC}$ and $w^{EE}$ to have an average of $1$, $2$, $4$ and $6$, respectively. Thanks to the powerful network infrastructure equipped in cloud centres, the intra-cloud communication should be the cheapest among communication directions. The download cost should be cheaper than the upload cost and hence the cost is set to reflect this pattern. Finally, due to the limited network capacity between edge devices, it is natural to possess the highest communication cost. 

The ratio parameter $\lambda$ that balances between cost and profit is set to be $-8$ so that the algorithm can jointly minimise the cost and profit times this negative ratio. 

\ul{\textit{PECCO-MFI Parameter Setting}}. We set the default allocation upper bound $ub$ to be $1$, and the Moth-flame shape parameter $b$ to be $1$ to comply with $ub$. If the shape parameter $b$ is set to be too small, then the shape of the spiral will be very tight and it will never allocate tasks to some nodes. On the other hand, if $b$ is too large, then the spiral will be too wide and it may generate offloading strategy which does not make sense. The default number of search candidates, i.e., moths, is set to be $30$, and the number of iterations is set to be $100$ to make the algorithm efficient. The default value of the threshold of the lifetime parameter $\tau$ is set to be $0.8$. 

\textbf{Experimental Setup and Hardware Configuration}. We compare the \emph{PECCO-MFI} algorithm with two edge-cloud computation offloading algorithms, including the LARAC algorithm, which traverses the shortest path during communication and optimises the computation cost, as well as GREEDY, which allocates each task to the side that will yield lower objective value, then greedily select the node which has the shortest distance from the initial location of the task. \rw{The effectiveness of the \emph{PECCO-MFI} is also compared with $9$ swarm-based optimisers, }including Bat Algorithm (BAT) \cite{yang2012bat}, Sine Cosine Algorithm (SCA) \cite{mirjalili2016sca}, Whale Optimisation Algorithm (WOA) \cite{mirjalili2016whale}, Cuckoo Search Algorithm (CS) \cite{yang2010engineering}, Firefly Algorithm (FFA) \cite{yang2013firefly}, Particle Swarm Optimisation (PSO) \cite{kennedy1995particle}, Grey Wolf Optimisation (GWO) \cite{2014Grey}, Differential Evolution (DE) \cite{price2013differential} and the original Moth-flame Optimisation Algorithm (MFO) \cite{moth_flame_optimisation_mirjalili2015moth}. To make the experimental results concrete, all experiments are repeated $10$ times and the average results are reported. 

We implement the method using python $3.8$, and conduct all experiments on a server equipped with Intel Core i9 9900K CPU and 32GB of memory. 

\begin{table*}[!ht]
  \centering
  \begin{tabular}{c|cccccc}
  \hline
  \backslashbox{Value}{Method} & LARAC & Greedy & BAT & SCA & WOA & CS \\ \hline
  Overall Objective & -19253.93 & -22615.6 & -41317.38 & -41496.97 & -42489.74 & -44088.8 \\
  Profit & 2650.29 & 3054.72 & 5388.28 & 5407.45 & 5531.59 & 5737.87 \\
  Cost & 1948.38 & 1822.16 & 1788.83 & \textbf{1762.64} & 1762.98 & 1814.15 \\
  Profit/Cost Ratio & 1.36 & 1.67 & 3.02 & 3.07 & 3.14 & 3.17 \\ 
  \hline
  \backslashbox{Value}{Method} & FFA & PSO & GWO & DE & MFO & MFI \\ \hline
  Overall Objective & -44741.63 & -45546.13 & -46494.95 & -46598.59 & -45953.91 & \textbf{-48069.48} \\
  Profit & 5814.99 & 5915.65 & 6039.85 & 6050.26 & 5969.91 & \textbf{6229.31} \\
  Cost & 1778.32 & 1779.07 & 1823.88 & 1803.52 & 1805.37 & 1765.0 \\
  Profit/Cost Ratio & 3.28 & 3.34 & 3.32 & 3.36 & 3.32 & \textbf{3.53} \\
  \hline
  \end{tabular}
  \caption{Objective value, profit, cost and the profit-cost ratio of different algorithms. Note the MFO stands for the original MFO algorithm, while the MFI stands for the improved \emph{PECCO-MFI} algorithm. }
  \label{tab:objective_profit_cost_pcratio}
\end{table*}

\subsection{Comparison of Objective Optimisation between Algorithm}
\label{sec:comparison_of_objective_optimisation_between_algorithm}

\begin{figure}[!ht]
  \begin{center}
    \includegraphics[width=0.26\textwidth]{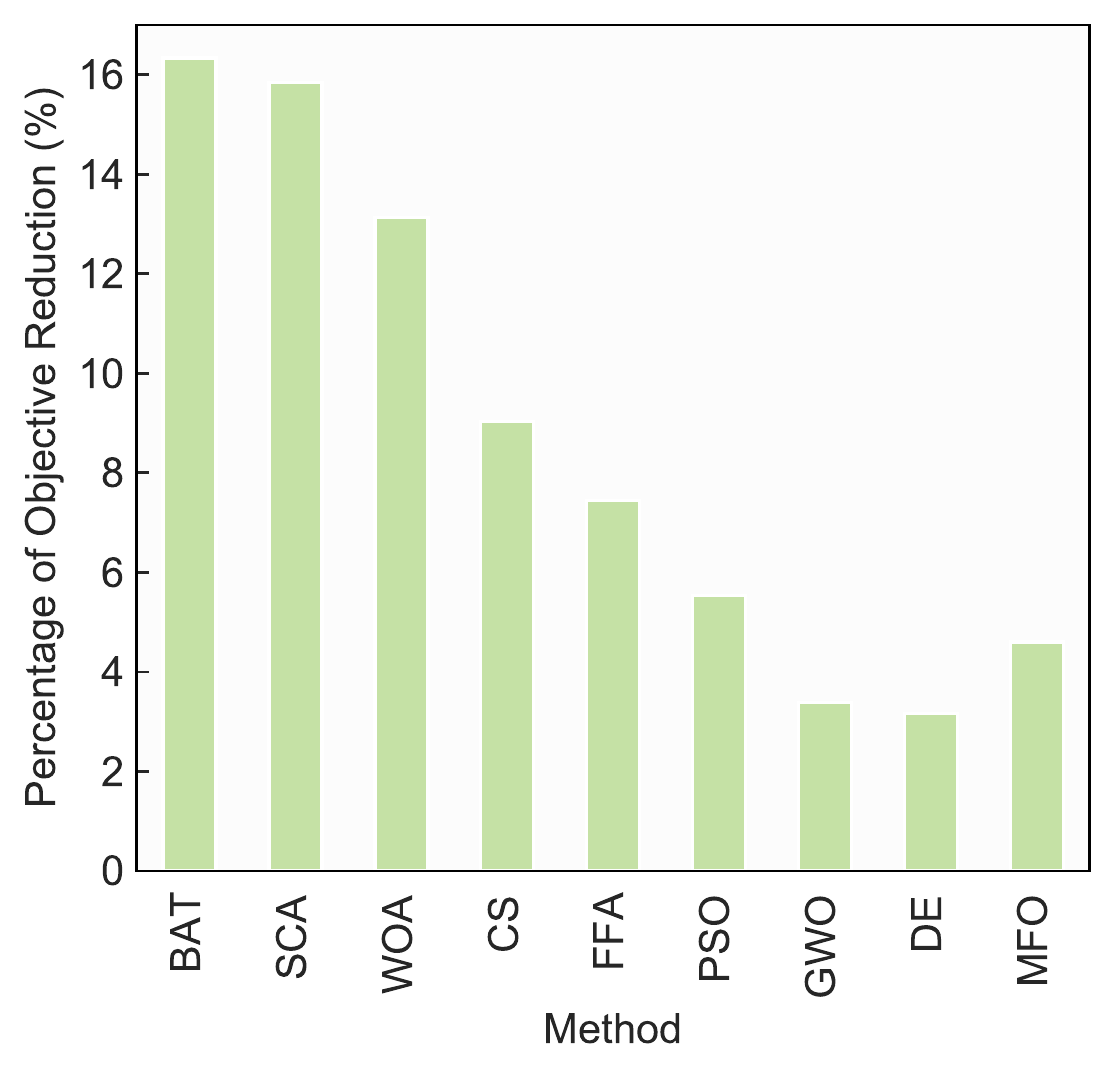}\\
    \caption{The percentage of objective value reduction achieved by the \emph{PECCO-MFI} algorithm compared with other methods. Since the objective value reduction yielded by the \emph{PECCO-MFI} is $149.7\%$ and $112.6\%$ compared with LARAC and GREEDY, respectively. Therefore, for better visualisation, these two methods are omitted from the plot. }
    \label{fig:figure_objective_reduction}
  \end{center}
\end{figure}

To verify the effectiveness of the \emph{PECCO-MFI} algorithm in terms of objective optimisation, the objective results of the \emph{PECCO-MFI} and $11$ compared methods are listed in Table \ref{tab:objective_profit_cost_pcratio}. As we can notice, the proposed \emph{PECCO-MFI} algorithm achieves the lowest objective value among all compared methods. As we can observe from Figure \ref{fig:figure_objective_reduction}, the \emph{PECCO-MFI} algorithm produces a $4.6\%$ and $3.16\%$ objective value reduction compared with the original Moth-flame Optimisation algorithm and the best-performed Differential Evolution algorithm when tackling this optimisation problem, demonstrating the effectiveness of the \emph{PECCO-MFI} algorithm. The significant performance improvement achieved by the \emph{PECCO-MFI} over the original Moth-flame Optimiser also verifies the effectiveness of improvements we made on the MFO. 

\begin{figure}[!ht]
  \begin{center}
    \includegraphics[width=0.5\textwidth]{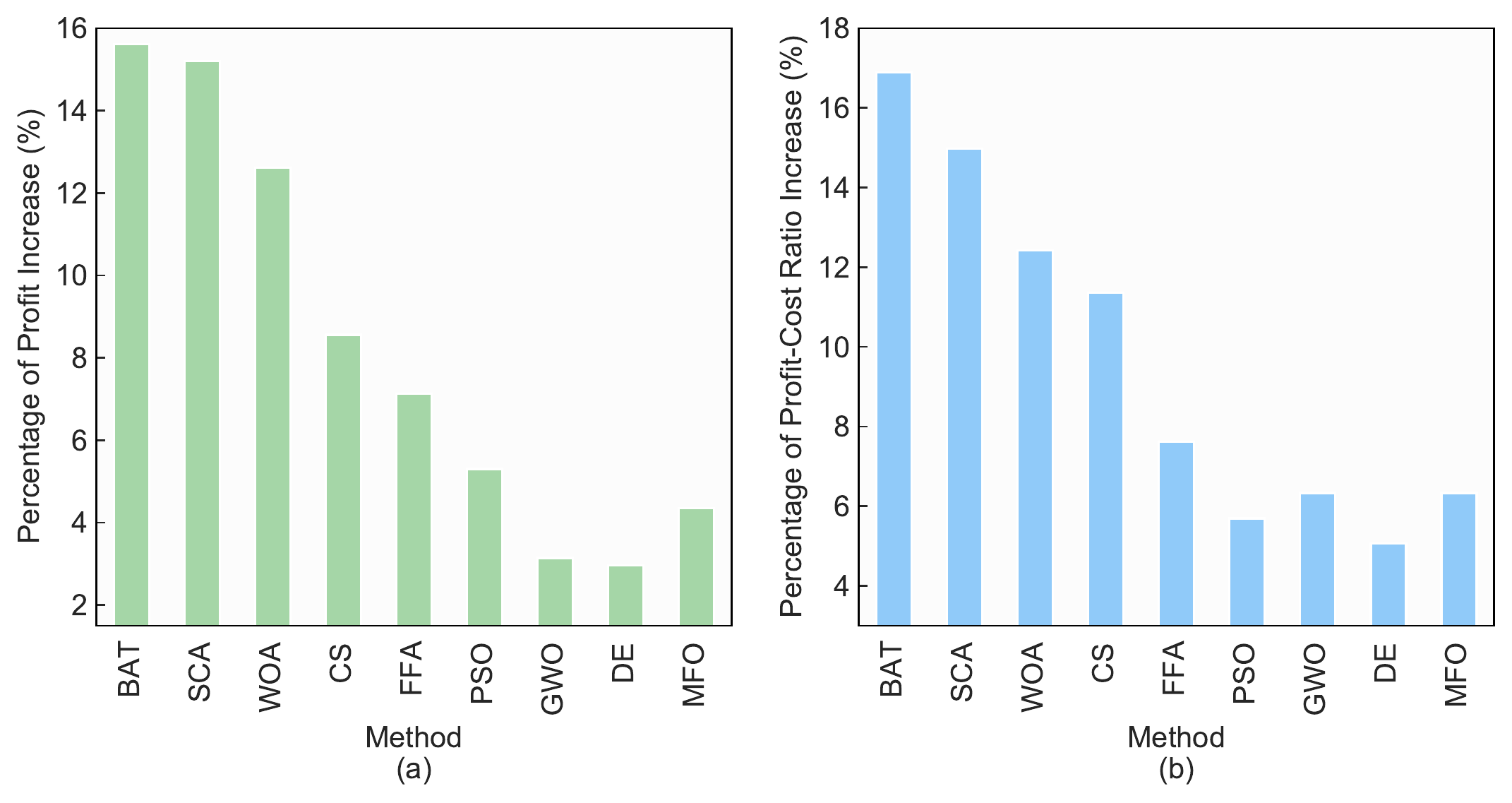}\\
    \caption{The percentage of increase on profit and profit-cost ratio achieved by the \emph{PECCO-MFI} algorithm compared with other methods are presented in sub-figure (a) and (b), respectively. Since the profit achieved by the \emph{PECCO-MFI} is $135\%$ and $103.9\%$ higher than LARAC and GREEDY, respectively, and the profit-cost ratio is $159.6\%$ and $111.4\%$ higher than LARAC and GREEDY, respectively. Therefore, for better visualisation, these two methods are omitted from plots. }
    \label{fig:figure_profit_pcratio_increase}
  \end{center}
\end{figure}

\vspace{0cm}

\subsection{Comparison of Profit and Cost between Algorithms}
\label{sec:comparison_of_profit_and_cost_between_algorithms}

After investigating the total objective, we now look at the profit and cost component. As indicated in Table \ref{tab:objective_profit_cost_pcratio} and Figure \ref{fig:figure_profit_pcratio_increase}(a), the \emph{PECCO-MFI} algorithm achieves the highest profit among all comparing methods. Specifically, the profit achieved is $4.35\%$ and $2.96\%$ higher than the original MFO algorithm, as well as the best-performed Differential Evolution, respectively. Besides, the \emph{PECCO-MFI} achieves the second lowest cost during computation offloading, which is only $0.1\%$ higher than the SCA algorithm, who has the lowest cost. However, when it comes to the profit-cost ratio, the \emph{PECCO-MFI} yields the best performance. As indicated in Figure \ref{fig:figure_profit_pcratio_increase}(b), the \emph{PECCO-MFI} algorithm achieves significant profit-cost ratio boost compared with all other methods. The higher the profit-cost ratio is, the more profit will be yielded by spending one unit of cost, i.e., the computation offloading is wiser as it can achieve higher profit by spending unit amount of cost. According to Figure \ref{fig:figure_profit_pcratio_increase}(b), the \emph{PECCO-MFI} has a profit-cost ratio that is $6.33\%$ and $5.06\%$ superior than the original MFO and Differential Evolution counterparts, respectively, which demonstrates the effectiveness of the \emph{PECCO-MFI} in terms of profit and cost-oriented offloading optimisation. It also indicates that the \emph{PECCO-MFI} can draw a computation offloading strategy that can utilise cost wisely to produce excellent profit. 

\begin{table*}[!ht]
  \centering
  \begin{tabular}{c|cccccc}
  \hline
  \backslashbox{Value}{Method} & LARAC & Greedy & BAT & SCA & WOA & CS \\ \hline
  \#Allocation & 64.9 & 92.9 & 175.3 & 170.7 & 177.2 & 176.7 \\
  Profit/Allocation Ratio & \textbf{40.96} & 31.69 & 30.74 & 31.68 & 31.22 & 32.48 \\ 
  Cost/Allocation Ratio & 30.02 & 19.61 & 10.20 & 10.33 & 9.95 & 10.27 \\
  \hline
  \backslashbox{Value}{Method} & FFA & PSO & GWO & DE & MFO & MFI \\ \hline
  \#Allocation & 178.6 & 179.2 & 179.3 & 177.6 & 179.1 & \textbf{179.3} \\
  Profit/Allocation Ratio & 32.57 & 33.69 & 33.66 & 34.06 & 33.33 & 34.75 \\
  Cost/Allocation Ratio & 9.96 & 9.93 & 10.17 & 10.15 & 10.08 & \textbf{9.84} \\
  \hline
  \end{tabular}
  \caption{Number of computation tasks being allocated, the profit-allocation ratio and the cost-allocation ratio of different algorithms. }
  \label{tab:profit_allocation_paratio}
\end{table*}

\vspace{0cm}

\subsection{Comparison of Task Allocation between Algorithms}
\label{sec:comparison_of_profit_and_task_completion_between_algorithms}

We now focus on task allocation done by different algorithms. If offloading in an unwise manner, some computing nodes will be overloaded, causing some tasks failed to be allocated. As we can see from Table \ref{tab:profit_allocation_paratio}, the proposed \emph{PECCO-MFI} algorithm achieves the highest task allocation number. A high number of tasks being allocated after computation offloading indicates that the \emph{PECCO-MFI} algorithm can offload tasks wisely without causing severe overloading. Hence, most number of tasks can be successfully allocated and completed instead of being stuck on overloaded computing nodes. 

\begin{figure}[!ht]
  \begin{center}
    \includegraphics[width=0.5\textwidth]{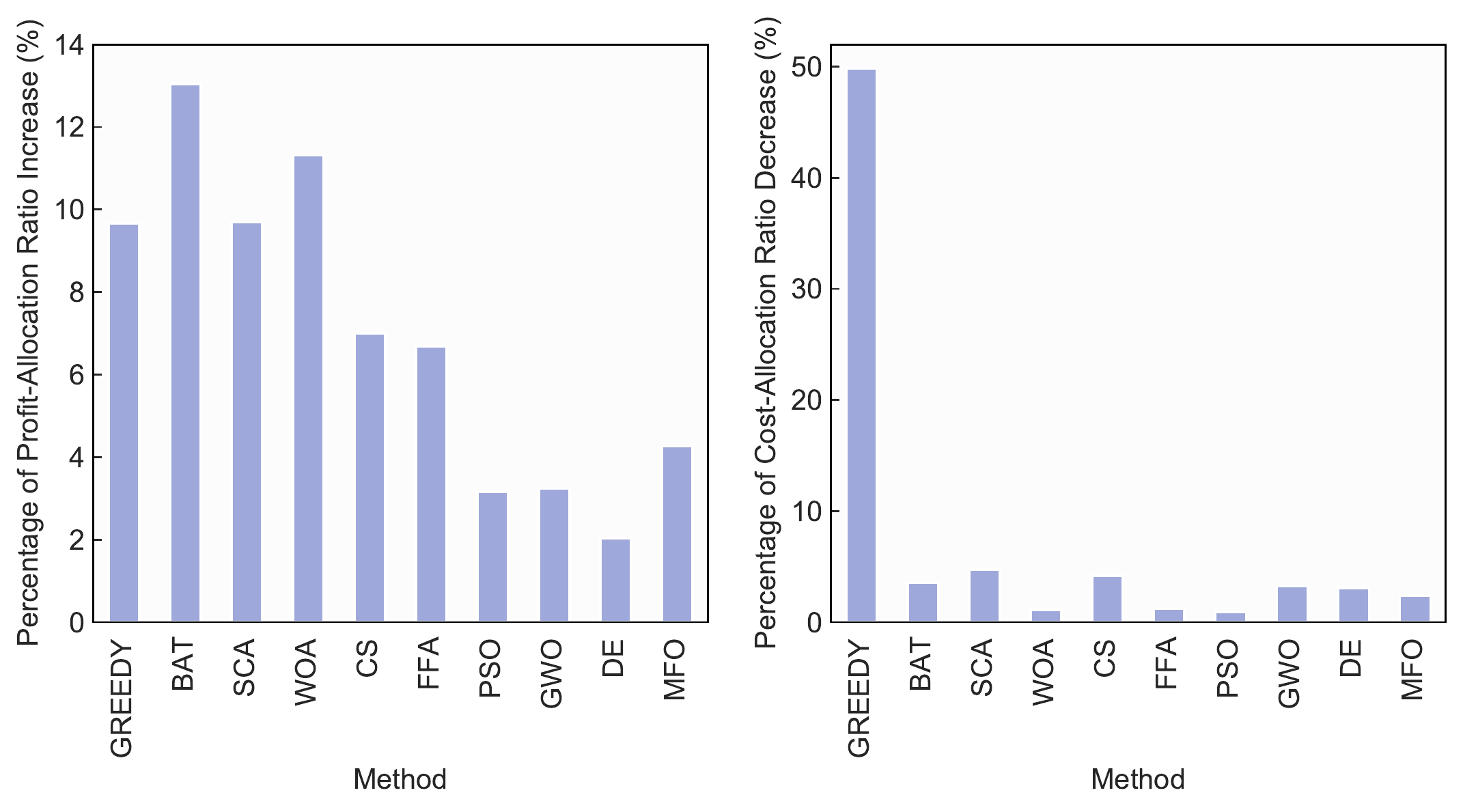}\\
    \caption{The percentage of profit-allocation ratio increase and cost-allocation ratio decrease achieved by the \emph{PECCO-MFI} algorithm compared with other methods are shown in (a) and (b), respectively. The extreme case LARAC is omitted for better visualisation. }
    \label{fig:figure_paratio_increase}
  \end{center}
\end{figure}

\vspace{0cm}

In terms of the profit-allocation ratio, both Table \ref{tab:profit_allocation_paratio} and Figure \ref{fig:figure_paratio_increase} show that except for the extreme case LARAC due to poor task allocation, the \emph{PECCO-MFI} algorithm produces the highest profit-allocation ratio. \rw{The higher the profit-allocation ratio is, the more profit will be yielded by completing each task. }As visualised in Figure \ref{fig:figure_paratio_increase}(a), the \emph{PECCO-MFI} algorithm achieves $4.26\%$ and $2.03\%$ higher profit-allocation ratio compared with the original MFO and the best-performed Differential Evolution, respectively. Besides, according to Table \ref{tab:profit_allocation_paratio}, the \emph{PECCO-MFI} yields the lowest cost-allocation ratio, which indicates the \emph{PECCO-MFI} causes the lowest cost to satisfy each task, demonstrating its cost-effectiveness. Hence, by achieving the highest number of task allocation, a high profit-allocation ratio and a low cost-allocation ratio, the effectiveness of the offloading strategy yielded by the \emph{PECCO-MFI} algorithm is verified. 

\begin{table*}[!ht]
  \centering
  \begin{tabular}{c|cccccc}
  \hline
  \backslashbox{Value}{Method} & LARAC & Greedy & BAT & SCA & WOA & CS \\ \hline
  Utilisation & \textcolor{grey}{121\%} & \textcolor{grey}{112\%} & 95\% & 92\% & 95\% & 95\% \\
  Profit/Utilisation Ratio & 22.12 & 32.51 & 56.97 & 59.02 & 57.95 & 60.28 \\ 
  Cost/Utilisation Ratio & \textbf{16.10} & 16.27 & 18.83 & 19.16 & 18.56 & 19.10 \\
  \hline
  \backslashbox{Value}{Method} & FFA & PSO & GWO & DE & MFO & MFI \\ \hline
  Utilisation & 95\% & 96\% & 95\% & 94\% & 96\% & 95\% \\
  Profit/Utilisation Ratio & 60.82 & 61.89 & 63.23 & 64.02 & 62.45 & \textbf{65.2} \\
  Cost/Utilisation Ratio & 18.72 & 18.53 & 19.20 & 19.19 & 18.81 & 18.58 \\
  \hline
  \end{tabular}
  \caption{Average computing node workload utilisation, the profit-utilisation ratio and the cost-utilisation ratio of different algorithms. }
  \label{tab:profit_utilisation_puratio}
\end{table*}

\vspace{0cm}

\subsection{Comparison of Resource Utilisation between Algorithms}
\label{sec:comparison_of_profit_and_computing_node_utilisation_between_algorithms}

Finally, the computing node workload resource utilisation, the profit-utilisation and cost-utilisation ratio are indicated in Table \ref{tab:profit_utilisation_puratio}. As we can observe, except the LARAC and GREEDY algorithm which overloads some computing nodes, all other methods produce offloading strategy that is free from overloading. 

\begin{figure}[!ht]
  \begin{center}
    \includegraphics[width=0.5\textwidth]{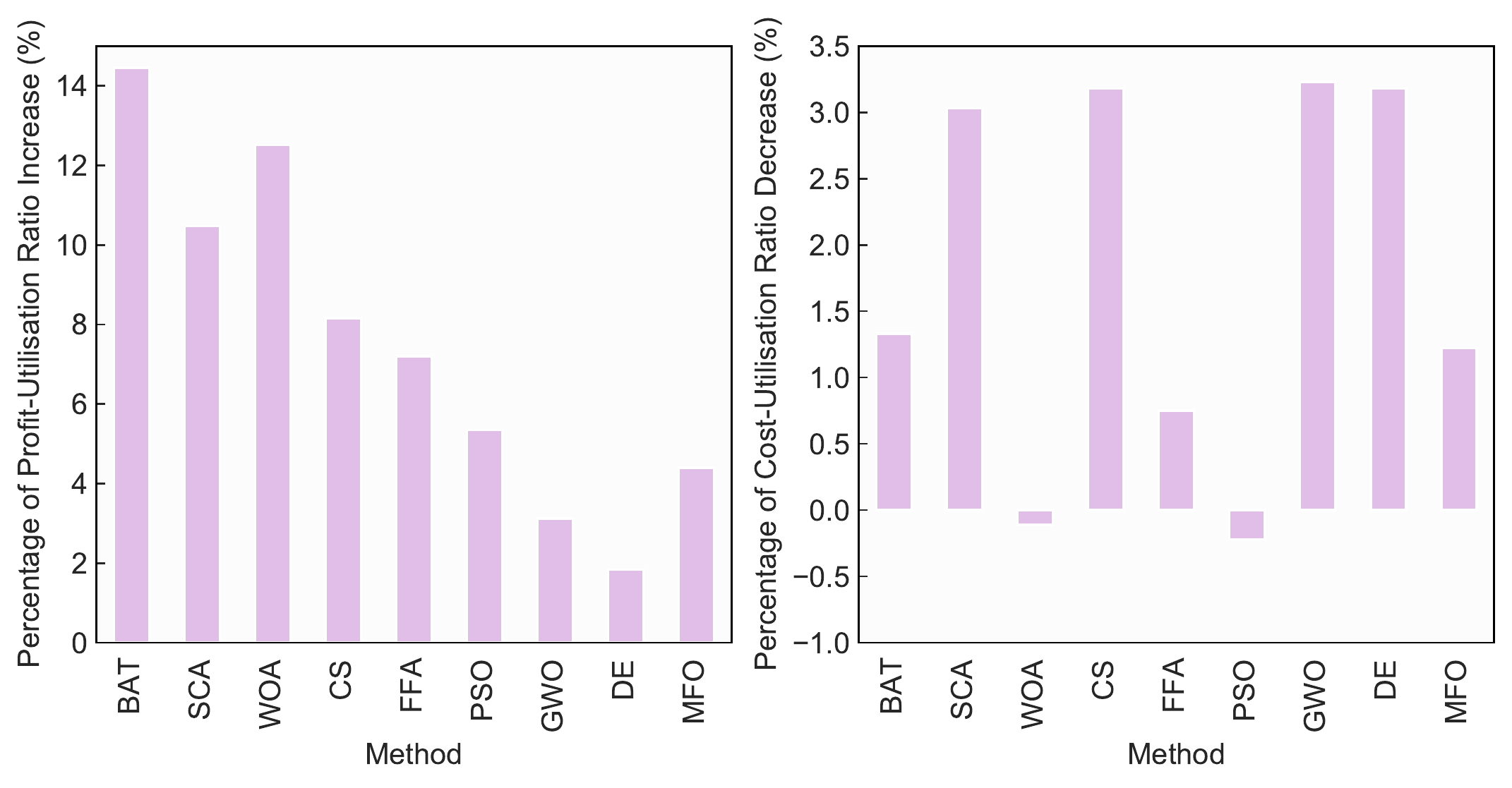}\\
    \caption{The percentage of profit-utilisation ratio increase and cost-utilisation ratio decrease achieved by the \emph{PECCO-MFI} algorithm compared with other methods are shown in (a) and (b), respectively. The extreme cases LARAC and GREEDY are omitted from the plot. }
    \label{fig:figure_puratio_increase}
  \end{center}
\end{figure}

As we can see from Figure \ref{fig:figure_puratio_increase}, the profit-utilisation ratio produced by the \emph{PECCO-MFI} algorithm is significantly higher than all other compared methods. Specifically, the \emph{PECCO-MFI} algorithm achieves a $4.4\%$ and $1.8\%$ increase in terms of profit-utilisation ratio compared with the original MFO and Differential Evolution, respectively. The higher the profit-utilisation ratio is, the more profit will be yielded by utilising one unit of computing node resource. \rw{On the other hand, the \emph{PECCO-MFI} yields a relatively low cost-utilisation ratio, which means the algorithm will not incur a high cost by utilising one unit of computation resource. }Hence, the excellent profit-utilisation and cost-utilisation ratio indicate the effectiveness of the \emph{PECCO-MFI} algorithm, i.e., producing a computation offloading strategy that can utilise computation resource wisely to achieve a high profit and a low cost, without overloading any computing nodes. 

\vspace{-1cm}

%%%%%%%%%%%%%%%%%%
%%% Conclusion %%%
%%%%%%%%%%%%%%%%%%

\section{Conclusion}
\label{sec:section_conclusion}

In this paper, we propose a profit and cost-oriented edge-cloud computation offloading model \emph{PECCO} which jointly considers the heterogeneous communication and computation cost, as well as the profit yielded after computation offloading. \rw{An improved Moth-flame Optimisation algorithm with three improvements is proposed which addresses several deficiencies of the original MFO and is then integrated to produce an optimised edge-cloud computation offloading strategy, }forming the \emph{PECCO-MFI} algorithm. Comprehensive experiments are conducted and the \emph{PECCO-MFI} algorithm is compared with several other baseline methods to testify to the effectiveness of the \emph{PECCO-MFI} algorithm when optimising the edge-cloud computation offloading model, \rw{as well as the effectiveness of the improvements made over the original MFO. }

\section*{Acknowledgement}
\label{sec:section_acknowledgement}

This work is supported in part by Key-Area Research and Development Program of Guangdong Province (2020B010164002) and Zhejiang Provincial Natural Science Foundation of China (LZ22F020002). 

\section*{Data Availability Statement}
\label{sec:section_data_availability_statement}

The data that support the findings of this study are available from the corresponding author upon reasonable request. 

\begin{spacing}{0.85}
\bibliography{PECCO-MF}

\begin{thebibliography}{10}
\providecommand \doibase [0]{http://dx.doi.org/}%

\bibitem{rapid_prevalence_smart_devices_shafique2020internet}
Shafique K, Khawaja BA, Sabir F, Qazi S, Mustaqim M. Internet of things (IoT)
  for next-generation smart systems: A review of current challenges, future
  trends and prospects for emerging 5G-IoT scenarios. {\it Ieee Access}
  2020\string; 8\string: 23022--23040.

\bibitem{8264678}
Shen S, Huang L, Zhou H, Yu S, Fan E, Cao Q. Multistage Signaling Game-Based
  Optimal Detection Strategies for Suppressing Malware Diffusion in
  Fog-Cloud-Based IoT Networks. {\it IEEE Internet of Things Journal}
  2018\string; 5(2)\string: 1043-1054.
\newblock \href {\doibase 10.1109/JIOT.2018.2795549} {doi:
  10.1109/JIOT.2018.2795549}

\bibitem{iot_app}
Zhang K, Tian J, Xiao H, Zhao Y, Zhao W, Chen J. A Numerical Splitting and
  Adaptive Privacy Budget Allocation Based LDP Mechanism for Privacy
  Preservation in Blockchain-Powered IoT. {\it IEEE Internet of Things Journal}
  2022\string: 1-1.
\newblock \href {\doibase 10.1109/JIOT.2022.3145845} {doi:
  10.1109/JIOT.2022.3145845}

\bibitem{9698094}
Li T, Wang H, He D, Yu J. Blockchain-based Privacy-preserving and Rewarding
  Private Data Sharing for IoT. {\it IEEE Internet of Things Journal}
  2022\string: 1-1.
\newblock \href {\doibase 10.1109/JIOT.2022.3147925} {doi:
  10.1109/JIOT.2022.3147925}

\bibitem{wu_theta_join}
Wu J, Wang Y, Fan X, Ye K, Xu C. Toward fast theta-join: A prefiltering and
  amalgamated partitioning approach. {\it Concurrency and Computation: Practice
  and Experience}\string; n/a(n/a)\string: e6996.
\newblock \href {\doibase https://doi.org/10.1002/cpe.6996} {doi:
  https://doi.org/10.1002/cpe.6996}

\bibitem{huge_data_generated_marjani2017big}
Marjani M, Nasaruddin F, Gani A, et al. Big IoT data analytics: architecture,
  opportunities, and open research challenges. {\it IEEE Access} 2017\string;
  5\string: 5247--5261.

\bibitem{li2021self_big_data}
Li M, Wu J, Dai J, et al. A self-contained and self-explanatory DNA storage
  system. {\it Scientific Reports} 2021\string; 11(1)\string: 1--15.

\bibitem{iot_limited_computation_shakarami2020survey}
Shakarami A, Ghobaei-Arani M, Masdari M, Hosseinzadeh M. A survey on the
  computation offloading approaches in mobile edge/cloud computing environment:
  a stochastic-based perspective. {\it Journal of Grid Computing} 2020\string;
  18(4)\string: 639--671.

\bibitem{9691460}
Li Q, Zhang Q, Huang H, Zhang W, Chen W, Wang H. Secure, Efficient and Weighted
  Access Control for Cloud-assisted Industrial IoT. {\it IEEE Internet of
  Things Journal} 2022\string: 1-1.
\newblock \href {\doibase 10.1109/JIOT.2022.3146197} {doi:
  10.1109/JIOT.2022.3146197}

\bibitem{SHEN2022103140}
Shen Y, Shen S, Wu Z, Zhou H, Yu S. Signaling game-based availability
  assessment for edge computing-assisted IoT systems with malware
  dissemination. {\it Journal of Information Security and Applications}
  2022\string; 66\string: 103140.
\newblock \href {\doibase https://doi.org/10.1016/j.jisa.2022.103140} {doi:
  https://doi.org/10.1016/j.jisa.2022.103140}

\bibitem{ar_ren2019edge}
Ren J, He Y, Huang G, Yu G, Cai Y, Zhang Z. An edge-computing based
  architecture for mobile augmented reality. {\it IEEE Network} 2019\string;
  33(4)\string: 162--169.

\bibitem{vr_zhang2017towards}
Zhang W, Chen J, Zhang Y, Raychaudhuri D. Towards efficient edge cloud
  augmentation for virtual reality mmogs. In: SEC '17. Association for
  Computing Machinery; 2017\string: 1--14.

\bibitem{burden_on_cloud_centre_mao2017mobile}
Mao Y, You C, Zhang J, Huang K, Letaief KB. Mobile edge computing: Survey and
  research outlook. {\it arXiv preprint arXiv:1701.01090} 2017.

\bibitem{burden_on_cloud_centre_yu2017survey}
Yu W, Liang F, He X, et al. A survey on the edge computing for the Internet of
  Things. {\it IEEE access} 2017\string; 6\string: 6900--6919.

\bibitem{burden_on_cloud_centre_shi2016edge}
Shi W, Cao J, Zhang Q, Li Y, Xu L. Edge computing: Vision and challenges. {\it
  IEEE internet of things journal} 2016\string; 3(5)\string: 637--646.

\bibitem{data_generated}
Zhao Y, Chen J. A Survey on Differential Privacy for Unstructured Data Content.
  {\it ACM Comput. Surv.} 2021.
\newblock Just Accepted\href {\doibase 10.1145/3490237} {doi: 10.1145/3490237}

\bibitem{burden_on_bandwidth_shi2016promise}
Shi W, Dustdar S. The promise of edge computing. {\it Computer} 2016\string;
  49(5)\string: 78--81.

\bibitem{edge_computing_survey_khan2019edge}
Khan WZ, Ahmed E, Hakak S, Yaqoob I, Ahmed A. Edge computing: A survey. {\it
  Future Generation Computer Systems} 2019\string; 97\string: 219--235.

\bibitem{heto_du2020algorithmics}
Du M, Wang Y, Ye K, Xu C. Algorithmics of cost-driven computation offloading in
  the edge-cloud environment. {\it IEEE Transactions on Computers} 2020\string;
  69(10)\string: 1519--1532.

\bibitem{computation_offloading_wang2019edge}
Wang J, Pan J, Esposito F, Calyam P, Yang Z, Mohapatra P. Edge cloud offloading
  algorithms: Issues, methods, and perspectives. {\it ACM Computing Surveys
  (CSUR)} 2019\string; 52(1)\string: 1--23.

\bibitem{qos_mach2017mobile}
Mach P, Becvar Z. Mobile edge computing: A survey on architecture and
  computation offloading. {\it IEEE Communications Surveys \& Tutorials}
  2017\string; 19(3)\string: 1628--1656.

\bibitem{homogeneous_communication_limited_factor_wang2017computational}
Wang W, Zhou W. Computational offloading with delay and capacity constraints in
  mobile edge. In: IEEE. ; 2017\string: 1--6.

\bibitem{limited_factor_li2001computation}
Li Z, Wang C, Xu R. Computation offloading to save energy on handheld devices:
  a partition scheme. In: CASES '01. Association for Computing Machinery;
  2001\string: 238--246.

\bibitem{larac_juttner2001lagrange}
Juttner A, Szviatovski B, M{\'e}cs I, Rajk{\'o} Z. Lagrange relaxation based
  method for the QoS routing problem. In: . 2. IEEE. ; 2001\string: 859--868.

\bibitem{homogeneous_communication_wu2016optimal}
Wu H, Knottenbelt W, Wolter K, Sun Y. An optimal offloading partitioning
  algorithm in mobile cloud computing. In: Springer. ; 2016\string: 311--328.

\bibitem{homogeneous_communication_dong2018computation}
Dong L, Wang F, Shan J. Computation offloading for mobile-edge computing with
  maximum flow minimum cut. In: CSAE '18. Association for Computing Machinery;
  2018\string: 1--5.

\bibitem{moth_flame_optimisation_mirjalili2015moth}
Mirjalili S. Moth-flame optimization algorithm: A novel nature-inspired
  heuristic paradigm. {\it Knowledge-based systems} 2015\string; 89\string:
  228--249.

\bibitem{9261459}
Liu J, Wang X, Shen S, Yue G, Yu S, Li M. A Bayesian Q-Learning Game for
  Dependable Task Offloading Against DDoS Attacks in Sensor Edge Cloud. {\it
  IEEE Internet of Things Journal} 2021\string; 8(9)\string: 7546-7561.
\newblock \href {\doibase 10.1109/JIOT.2020.3038554} {doi:
  10.1109/JIOT.2020.3038554}

\bibitem{SHI2018316}
Shi H, Liu S, Wu H, et al. Oscillatory Particle Swarm Optimizer. {\it Applied
  Soft Computing} 2018\string; 73\string: 316-327.
\newblock \href {\doibase https://doi.org/10.1016/j.asoc.2018.08.037} {doi:
  https://doi.org/10.1016/j.asoc.2018.08.037}

\bibitem{lai2020analysis}
Lai X, Zhou Y. Analysis of multiobjective evolutionary algorithms on the
  biobjective traveling salesman problem (1, 2). {\it Multimedia Tools and
  Applications} 2020\string; 79(41)\string: 30839--30860.

\bibitem{hill_climbing_davis1991bit}
Davis L. Bit-climbing, representational bias, and test suit design. In: ;
  1991\string: 18--23.

\bibitem{ils_lourencco2003iterated}
Louren{\c{c}}o HR, Martin OC, St{\"u}tzle T. Iterated local search. In:
  Springer.  2003 (pp. 320--353).

\bibitem{gradient_descent_ruder2016overview}
Ruder S. An overview of gradient descent optimization algorithms. {\it arXiv
  preprint arXiv:1609.04747} 2016.

\bibitem{goodfellow2016deep}
Goodfellow I, Bengio Y, Courville A. {\it Deep learning}.
\newblock MIT press .
\newblock 2016.

\bibitem{li2020simultaneous}
Li S, Xie B, Wu J, Zhao Y, Liu CH, Ding Z. Simultaneous Semantic Alignment
  Network for Heterogeneous Domain Adaptation. In: Association for Computing
  Machinery; 2020\string: 3866--3874.

\bibitem{swarm_keerthi2015survey}
Keerthi S, Ashwini K, Vijaykumar M. Survey paper on swarm intelligence. {\it
  International Journal of Computer Applications} 2015\string; 115(5).

\bibitem{kennedy1995particle}
Kennedy J, Eberhart R. Particle swarm optimization. In: . 4. IEEE. ;
  1995\string: 1942--1948.

\bibitem{yang2013firefly}
Yang XS, He X. Firefly algorithm: recent advances and applications. {\it
  International journal of swarm intelligence} 2013\string; 1(1)\string:
  36--50.

\bibitem{mirjalili2016whale}
Mirjalili S, Lewis A. The whale optimization algorithm. {\it Advances in
  engineering software} 2016\string; 95\string: 51--67.

\bibitem{2014Grey}
Sm A, Smm B, Al A. Grey Wolf Optimizer. {\it Advances in Engineering Software}
  2014\string: 46–61.

\bibitem{genetic_algorithms_framework_mirjalili2019genetic}
Mirjalili S. Genetic algorithm. In: Springer.  2019 (pp. 43--55).

\bibitem{huang2019dp_greedy}
Huang D, Fan X, Wang Y, He S, Xu C. DP\_Greedy: A Two-Phase Caching Algorithm
  for Mobile Cloud Services. In: IEEE. ; 2019\string: 1--10.

\bibitem{nsw_parking}
Transportation N. Commuter Carparks TfNSW Open Data Hub and Developer Portal.
  \url{https://opendata.transport.nsw.gov.au/dataset/commuter-carparks}; .
\newblock Accessed: 2021-07-01.

\bibitem{yang2012bat}
Yang XS, Gandomi AH. Bat algorithm: a novel approach for global engineering
  optimization. {\it Engineering computations} 2012.

\bibitem{mirjalili2016sca}
Mirjalili S. SCA: a sine cosine algorithm for solving optimization problems.
  {\it Knowledge-based systems} 2016\string; 96\string: 120--133.

\bibitem{yang2010engineering}
Yang XS, Deb S. Engineering optimisation by cuckoo search. {\it International
  Journal of Mathematical Modelling and Numerical Optimisation} 2010\string;
  1(4)\string: 330--343.

\bibitem{price2013differential}
Price KV. Differential evolution. In: Springer.  2013 (pp. 187--214).

\end{thebibliography}
\end{spacing}

\end{document}